\documentclass[aps,prl,reprint,groupedaddress,showpacs,amsmath,amssymb,twocolumn,floatfix]{revtex4-1}
\usepackage{graphicx}
\usepackage{bm}
\usepackage{color}
\usepackage[normalem]{ulem}

\usepackage{amsmath}
\usepackage{hyperref}
\usepackage{cleveref}
\usepackage{cancel}
\usepackage{url}
\usepackage{xcolor}

\begin{document}

\title{Bilayer Excitons in the Laughlin Fractional Quantum Hall State}

\author{Ron Q. Nguyen$^{1}$}
\thanks{These authors contributed equally to this work.}
\author{Naiyuan J. Zhang$^{1}$}
\thanks{These authors contributed equally to this work.}
\author{Navketan Khurana-Batra$^{1,2}$}
\thanks{These authors contributed equally to this work.}
\author{Sarah Alkidim$^{1}$}
\author{Xiaoxue Liu$^{1}$}
\author{Kenji Watanabe$^{3}$}
\author{Takashi Taniguchi$^{4}$}
\author{D. E. Feldman$^{1,2}$}
\author{J.I.A. Li$^{1,5}$}
\thanks{Email: jia$\_$li@brown.edu}

\affiliation{$^{1}$Department of Physics, Brown University, Providence, Rhode Island 02912, USA}
\affiliation{$^{2}$Brown Theoretical Physics Center, Brown University, Providence, Rhode Island 02912, USA}
\affiliation{$^{3}$Research Center for Electronic and Optical Materials, National Institute for Materials Science, 1-1 Namiki, Tsukuba 305-0044, Japan}
\affiliation{$^{4}$Research Center for Materials Nanoarchitectonics, National Institute for Materials Science,  1-1 Namiki, Tsukuba 305-0044, Japan}
\affiliation{$^{5}$Department of Physics, University of Texas at Austin, Austin, TX 78712, USA}

\date{\today}

\maketitle

\textbf{The Laughlin state embodies a universal class of fractional quantum Hall effects arising in two-dimensional electron systems subjected to strong perpendicular magnetic fields. 
Conventionally described by a single-component wavefunction, the Laughlin state features fractionally charged quasiparticles arising from correlations within one electron species. Here, we explore a novel physical situation by introducing inter-species Coulomb coupling between two intra-species Laughlin states in a quantum Hall graphene bilayer structure. Although quasiparticle excitations typically exhibit charge gaps of tens of Kelvin, we observe that this energy scale is significantly lowered through interlayer excitonic pairing between quasiparticles and quasiholes. Identified via transport measurements, these excitons belong to an unprecedented category of charge-neutral anyons, opening a new avenue for investigating exotic quantum statistics and phases of matter.
}

\begin{figure*}
\includegraphics[width=0.85\linewidth]{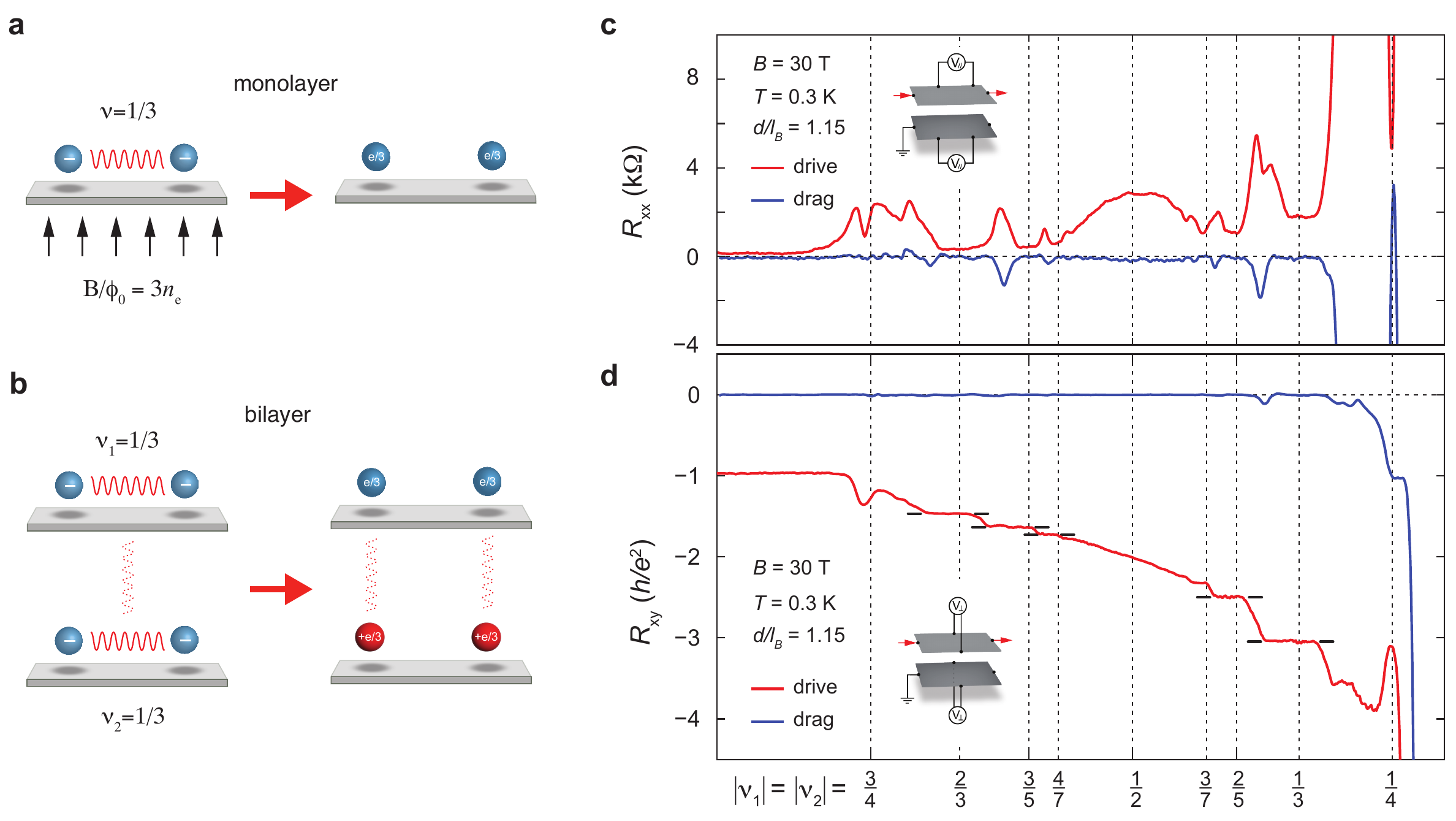}
\caption{\label{fig1}{\bf{Laughlin states in a quantum Hall graphene bilayer.}} 
(a) Schematic of a single-layer Laughlin state stabilized by intralayer Coulomb interactions. At Landau-level filling $\nu=1/3$, its quasiparticle excitations carry a fractional charge of $e/3$. 
(b) Schematic of two coupled Laughlin states in a quantum Hall bilayer, each stabilized by intralayer interactions. In this configuration, quasiparticles and quasiholes residing in opposite layers interact through weak interlayer Coulomb coupling. 
(c–d) Counterflow drag measurements as a function of Landau-level filling along the equal-density line $\nu_1 = \nu_2$. Data are acquired from a quantum Hall bilayer device patterned into a Hall-bar geometry. (c) Longitudinal and (d) transverse responses are shown for the drive and drag layers at $B = 30$~T and $T = 0.3$~K. Insets illustrate the corresponding measurement configurations.
}
\end{figure*}

\begin{figure*}
\includegraphics[width=0.85\linewidth]{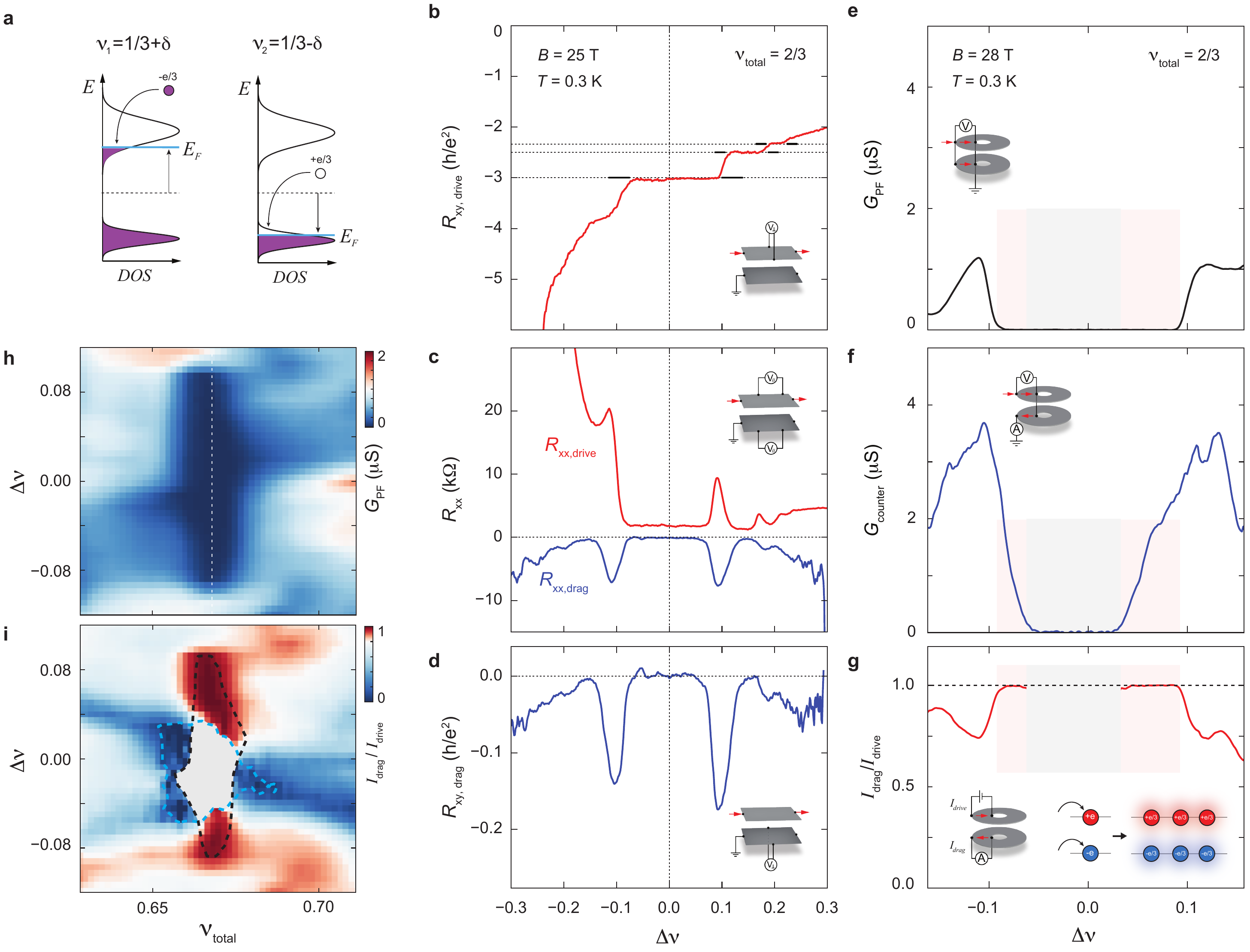}
\caption{\label{fig2} 
\textbf{Pairing between quasiparticles and quasiholes of the Laughlin state.} 
(a) Schematic illustrating how the population of bilayer excitons can be tuned by introducing layer-asymmetric doping. This is achieved by varying the Landau-level fillings of layers 1 and 2 by $\pm \delta$, thereby generating equal numbers of quasiparticles and quasiholes in opposite layers. 
(b) Hall resistance of the drive layer, $R_{xy,\mathrm{drive}}$, (c) longitudinal resistances of the drive and drag layers, $R_{xx,\mathrm{drive}}$ and $R_{xx,\mathrm{drag}}$, and (d) Hall resistance of the drag layer, $R_{xy,\mathrm{drag}}$, as functions of $\Delta\nu$ measured at total filling $\nu_{\mathrm{total}} = 2/3$. 
(e) Parallel-flow conductance $G_{\mathrm{PF}}$, (f) counterflow conductance $G_{\mathrm{counter}}$, and (g) drag current ratio $I_{\mathrm{drag}}/I_{\mathrm{drive}}$ as functions of $\Delta\nu$, measured at the same $\nu_{\mathrm{total}} = 2/3$. 
(h) Map of $G_{\mathrm{PF}}$ and (i) map of $I_{\mathrm{drag}}/I_{\mathrm{drive}}$ across the $\nu_{\mathrm{total}}$–$\Delta\nu$ plane near the Laughlin state at $\nu_1 = \nu_2 = 1/3$ and $B = 28$~T. 
Near $\Delta\nu = 0$, both $I_{\mathrm{drag}}$ and $I_{\mathrm{drive}}$ vanish, resulting in an ill-defined drag ratio, indicated by the gray-shaded regions. Panels (b-d) are measured in the Hall-bar-shaped device; panels (e-i) are measured in the Corbino-shaped device.}
\end{figure*}

\begin{figure*}
\includegraphics[width=0.85\linewidth]{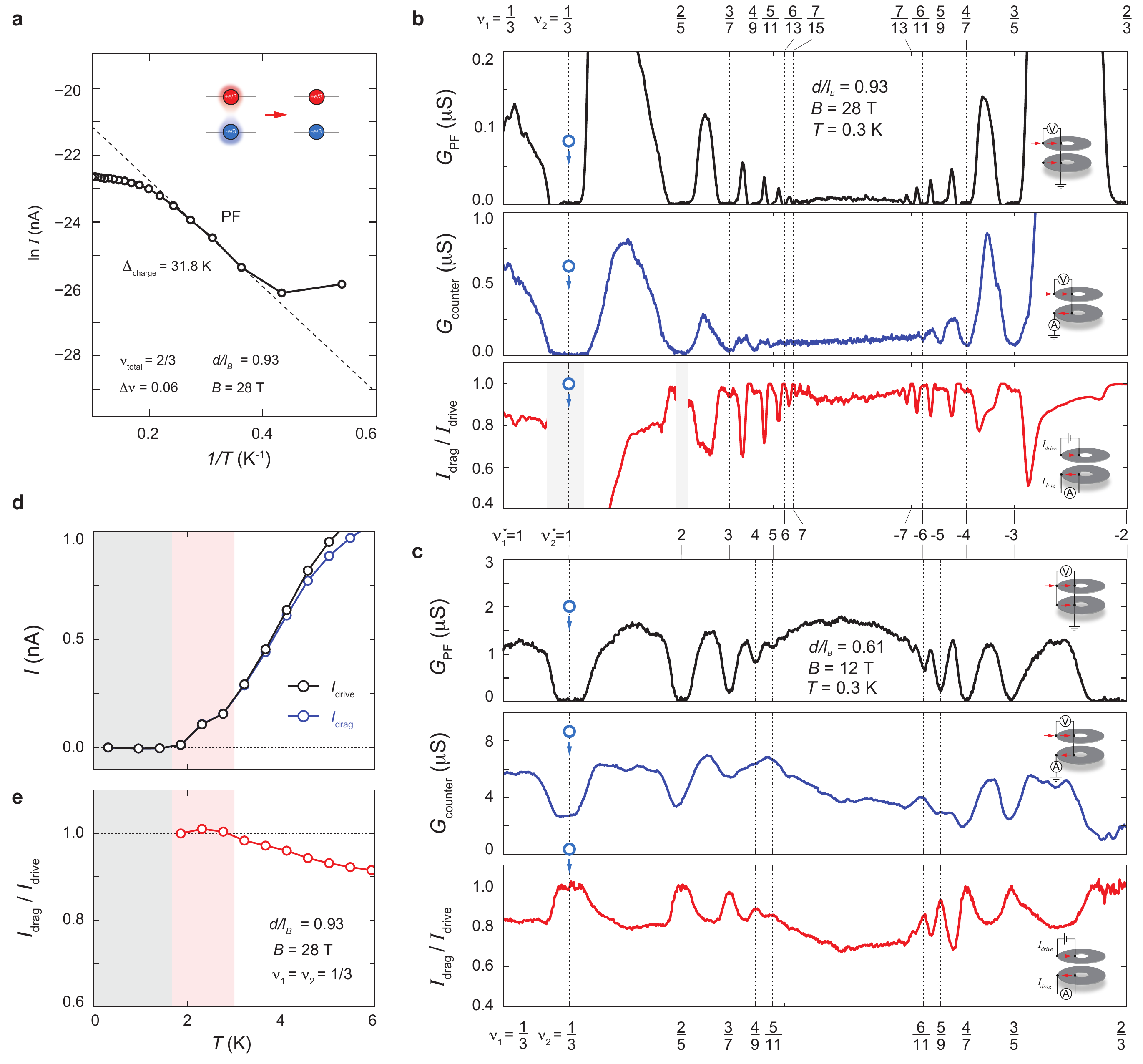}
\caption{\label{fig3} 
\textbf{Exciton binding energy.} 
(a) Arrhenius plot of the parallel-flow current at $\nu_{\mathrm{total}} = 2/3$ and $\Delta\nu = 0.06$. 
(b–c) Bulk transport measured at $\nu_1 = 1/3$ as a function of $\nu_2$ for (b) $B = 28$~T and (c) $B = 12$~T. 
The top, middle, and bottom panels show, respectively, the parallel-flow conductance $G_{\mathrm{PF}}$, counterflow conductance $G_{\mathrm{counter}}$, and drag current ratio $I_{\mathrm{drag}}/I_{\mathrm{drive}}$.  
(d–e) Temperature dependence of (d) drive and drag currents $I_{\mathrm{drive}}$ and $I_{\mathrm{drag}}$, and (e) the drag ratio $I_{\mathrm{drag}}/I_{\mathrm{drive}}$, measured at $\nu_1 = \nu_2 = 1/3$ and $B = 28$~T. 
The gray-shaded region marks the low-temperature regime where both currents vanish. 
The red-shaded region indicates the onset of simultaneous $I_{\mathrm{drive}}$ and $I_{\mathrm{drag}}$, corresponding to a perfect drag ratio. 
At higher temperatures, $I_{\mathrm{drive}}$ and $I_{\mathrm{drag}}$ bifurcate, and the drag ratio departs from unity. Data in this figure are measured in the Corbino-shaped device.
}
\end{figure*}

\begin{figure*}
\includegraphics[width=0.98\linewidth]{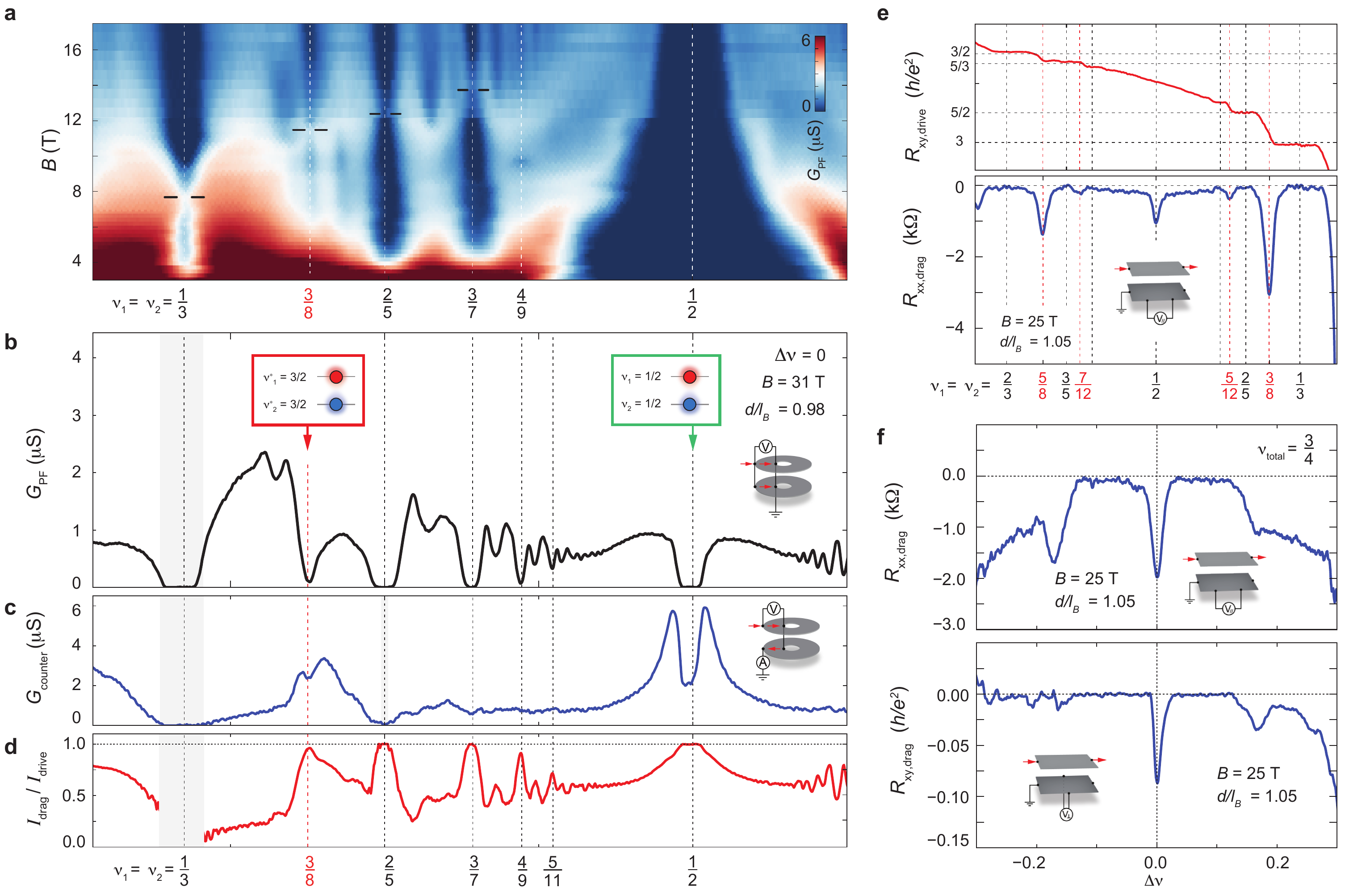}
\caption{\label{fig4}
\textbf{Exciton pairing at half-filled $\Lambda$-levels.} 
(a) Parallel-flow conductance $G_{\mathrm{PF}}$ as a function of $\nu_{\mathrm{total}}$ and $B$ along the equal-density line, revealing a sequence of magnetic-field–induced transitions in the FQH states, marked by horizontal black lines. 
In the high-field limit, the FQH states occur at integer and half-integer fillings of the $\Lambda$-levels, indicative of exotic exciton formation. 
(b) Parallel flow conductance $G_{\mathrm{PF}}$, (c) counterflow conductance $G_{\mathrm{counter}}$, and (d) drag current ratio $I_{\mathrm{drag}}/I_{\mathrm{drive}}$ as functions of $\nu_{\mathrm{total}}$, measured at $B = 31$~T. 
(e) Hall resistance on the drive layer $R_{xy,\mathrm{drive}}$ and longitudinal drag response $R_{xx,\mathrm{drag}}$ as a function of $\nu_1 = \nu_2$ along the equal-density line, measured in a Hall-bar device at $B = 25$~T. 
(f) Longitudinal and Hall drag response, $R_{xx,\mathrm{drag}}$ and $R_{xy,\mathrm{drag}}$, as functions of layer imbalance $\Delta\nu$, measured at $\nu_1 = \nu_2 = -3/8$ and $B = 25$~T. Panels (a-d) are measured in the Corbino-shaped device; panels (e-f) are measured in the Hall-bar-shaped device.
}
\end{figure*}

Quantum statistics form the cornerstone for understanding diverse quantum phases of matter in condensed matter physics. Fermionic statistics fundamentally describe electronic systems at low temperatures, underpinning the physics of Fermi liquids \cite{Luttinger1960fermi,Nozieres2018quantumliquids} and the quantum Hall effect \cite{Halperin2020fractional}. When fermions pair, they form composite particles which are bosons, enabling the emergence of low-temperature ground states characterized by Bose–Einstein condensation. Superconductivity, for instance, results from electrons near the Fermi surface undergoing Cooper pairing instabilities \cite{Cooper2011BCS}. Alternatively, excitonic superfluids represent a distinct class of bosonic condensate, resulting from Coulomb-driven electron-hole pairing, establishing an intriguing phenomenon that is actively pursued in condensed matter research \cite{Blatt1962exciton,Lozovik1975exciton,Lozovik1976exciton}.

While excitons are accessible in a wide range of materials, quantum Hall bilayers represent a paradigmatic solid-state platform for stabilizing excitonic condensates. In these bilayer structures, charge carriers reside in two closely spaced but electrically isolated two-dimensional layers. Coulomb attraction across the layers drives electron-hole pairing, stabilizing various emergent excitonic phases extensively studied both experimentally and theoretically \cite{Nandi2012exciton,Eisenstein2014,Li2017superfluid,Liu2017superfluid,Li2019pairing,Liu2022crossover,Eisenstein2004BEC,Kellogg2005,Zeng2023solid,Zhang2025fractionalexciton,Kellogg2003,Kellogg2004,Tutuc2004counterflow}. Among the most extensively examined phases, an excitonic condensate emerges at a total Landau level (LL) filling factor of one, accurately described by the (111) wavefunction \cite{Halperin1983,Wen1992}, representing an excitonic analogue of the integer quantum Hall effect state.

More recently, it was discovered that interlayer excitons can coexist with fractional quantum Hall states, indicating that excitonic phenomena are more ubiquitous within the quantum Hall regime than previously recognized. In the regime with strong interlayer correlation, fractional quantum Hall states can be described by two-component wavefunctions, notably the (333) state \cite{Halperin1983,Wen1992}, serving as the fractional analogue to the (111) state. Here, interlayer excitons naturally emerge as essential constituents of the ground-state construction \cite{Zhang2025fractionalexciton}.

In this work, we report a new pathway for creating interlayer excitons by introducing  Coulomb coupling between two Laughlin states. While the Laughlin state is traditionally regarded as a single-component state, we show that Coulomb coupling between two such states generates interlayer excitons at energies substantially lower than the charge gap. As shown in Fig.~\ref{fig1}a, quasiparticle excitation of the Laughlin state carries $1/3$ of an electron charge and obeys anyonic quantum statistics \cite{Tsui1982FQHE,halperin1984,Arovas1984fractional,review-FH,Picciotto1997,Saminadayar1997,Reznikov1999,Bartolomei2020fractional,Nakamura2020Anyon,
texp1,texp2,Willett2023interference}. We propose that interlayer excitons of two Laughlin states results from pairing between  fractionally charged quasiparticles and quasiholes (Fig.~\ref{fig1}b).

%\vspace{0.1 in}
%\noindent\textbf{Laughlin states in quantum Hall graphene bilayer}
%\vspace{0.1 in}

The phase space of a quantum Hall bilayer is defined by the Landau-level fillings of the two graphene layers, $\nu_1$ and $\nu_2$, which can be independently tuned by adjusting the voltages on the top and bottom gate electrodes. 
The strength of interlayer coupling is commonly parameterized by the ratio $d/\ell_B$, where $d$ is the interlayer separation and $\ell_B$ is the magnetic length. 
In this work, we report measurements from two comparable quantum Hall bilayer samples, each consisting of two monolayer graphene layers separated by an insulating barrier.
One device is patterned into a Hall-bar geometry~\cite{Li2017superfluid,Liu2017superfluid,Liu2022crossover,Eisenstein2014}  with $d=5.4$~nm, and the other into a Corbino geometry~\cite{Zeng2019,Li2019pairing,Zeng2023solid,Zhang2025fractionalexciton,Eisenstein2014,Nandi2012exciton} with $d=4.5$~nm. 
In the lowest Landau level (LL), monolayer graphene is electron–hole symmetric. We therefore focus on the lowest hole-doped LL in each layer, with $\nu_1$ and $\nu_2$ denoting the corresponding hole filling factors. 
Further details regarding device fabrication and measurement configurations are provided in the Methods section.

The Laughlin states emerge in the weak-coupling regime around $d/\ell_B \sim 1$, where each layer follows the conventional Jain sequence of fractional quantum Hall (FQH) states. 
Figures~\ref{fig1}c--d show counterflow drag measurements from a Hall-bar device as a function of Landau-level filling along the equal-density line, $\nu_1 = \nu_2$. 
These data reveal a series of Jain states at $\nu_1 = \nu_2 = N/(1 + 2N)$ with $N \in \mathbb{Z}$~\cite{Jain.03,Eisenstein1990FQHE}. 
In this regime, the bilayer hosts two decoupled copies of the Laughlin state at $\nu_1 = \nu_2 = 1/3$. 
Because each layer possesses a robust charge gap, no free charge carriers are available to support exciton pairing across the layers. 
Two observations confirm the absence of interlayer correlations: the drive layer exhibits a quantized Hall plateau at $R_{xy,\mathrm{drive}} = h/(\nu e^2)$, while the drag layer shows negligible signal in both longitudinal and Hall channels.

%In the above mentioned scenario, quasiparticles, quasiholes, and the resulting excitons are all thermal excitations. 
\vspace{0.1 in}
\noindent\textbf{Exciton pairing between Laughlin states}
%\vspace{0.1 in}

Starting from the pair of Laughlin states, interlayer excitons can be introduced by asymmetrically doping the graphene bilayer. 
As illustrated in Fig.~\ref{fig2}a, an equal population of quasiparticles and quasiholes is generated across the two layers by tuning the fillings to $\nu_1 = 1/3 + \delta$ and $\nu_2 = 1/3 - \delta$. 
This defines a vertical trajectory in the $\nu_{\mathrm{total}}-\Delta\nu$ phase space at constant total filling $\nu_{\mathrm{total}} = 2/3$, where the parameter $\delta$ directly quantifies the layer imbalance $\Delta\nu = 2\delta$.

The presence of bilayer excitons is revealed by characteristic signatures in the counterflow drag measurements. 
Figures~\ref{fig2}b--d show the drag responses from a Hall-bar device as a function of $\Delta\nu$ at $\nu_{\mathrm{total}} = 2/3$. 
As the bilayer is asymmetrically doped away from $\Delta\nu = 0$, the Hall response of the drive layer begins to deviate from the quantized Laughlin plateau around $\Delta\nu \approx -0.08$. 
Concurrently, pronounced peaks emerge in both the longitudinal and Hall drag resistances, $R_{xx,\mathrm{drag}}$ and $R_{xy,\mathrm{drag}}$, signaling the onset of strong interlayer correlations associated with exciton formation.

Definitive evidence of exciton pairing is obtained from bulk transport measurements using a Corbino-shaped sample. 
As shown in Figs.~\ref{fig2}e--g, the parallel-flow conductance $G_{\mathrm{PF}}$ vanishes within the range $-0.1 < \Delta\nu < +0.1$, highlighting a regime with a fully developed charge gap. 
This allows us to identify bilayer excitons through counterflow and drag measurements. 

Within this gapped regime, the counterflow conductance also vanishes near $\Delta\nu = 0$, as indicated by the gray-shaded region in Figs.~\ref{fig2}e--g. 
Because both the drive and drag currents are zero, the drag ratio is ill-defined in this region. 
As $\Delta\nu$ is tuned away from zero, the counterflow conductance rises from zero, giving way to a regime where finite counterflow conductance coexists with vanishing parallel-flow conductance, denoted by the red-shaded area in Figs.~\ref{fig2}e--g. 
Most remarkably, drag measurements reveal a perfect drag response in this regime, with a drag ratio of unity ($I_{\mathrm{drag}}/I_{\mathrm{drive}} = 1$). 
Together, these transport responses provide compelling evidence for the formation of interlayer excitons. 
Remarkably, in this regime the only charge-carrying entities are excitons, which sustain counterflowing currents while no free charge transport occurs in the parallel-flow channel.

Figures~\ref{fig2}h--i present bulk transport data around $\nu_1 = \nu_2 = 1/3$. 
The regime with a robust charge gap is identified by vanishing parallel-flow conductance, represented by dark blue in Fig.~\ref{fig2}h. 
While this region extends over a range of approximately $0.20$ along the $\Delta\nu$ axis, it spans only about $0.02$ along the $\nu_{\mathrm{total}}$ axis. 
This pronounced anisotropy suggests that the charge gap is stabilized by exciton formation along the direction of varying $\Delta\nu$ at constant $\nu_{\mathrm{total}}$. 
Indeed, a perfect drag response—represented by red in Fig.~\ref{fig2}i—is observed near the tips of the elongated charge-gap region, where finite $\Delta\nu$ is expected to induce interlayer excitons. 
In contrast, varying $\nu_{\mathrm{total}}$ at $\Delta\nu = 0$ closes the charge gap and restores finite parallel-flow conductance.

\vspace{0.1 in}
\noindent\textbf{Thermal excitation and exciton binding energy}
%\vspace{0.1 in}

The formation of excitons is associated with a characteristic binding energy, which is characterized in the following.  
Figure~\ref{fig3}a shows the Arrhenius plot of the parallel-flow conductance measured at $\nu_{\mathrm{total}} = 2/3$ and $\Delta\nu = 0.06$, where a perfect drag response is observed at low temperature. 
The activated behavior reflects the energy required to generate free charge carriers from a ground state containing excitons. 
Within this interpretation, the energy gap extracted from the Arrhenius plot provides a lower bound for the exciton binding energy.

Figures~\ref{fig3}b--c present transport measurements obtained by sweeping $\nu_2$ while holding $\nu_1$ fixed at $1/3$. At $B = 28$~T, the parallel-flow conductance vanishes at $\nu_2 = N/(1 + 2N)$ (top panel of Fig.~\ref{fig3}b), consistent with the robust charge gaps of the Jain sequence. Owing to the large charge gap at $\nu_1 = \nu_2 = 1/3$, excitons are absent, as indicated by the vanishing counterflow conductance and the ill-defined drag ratio.

Remarkably, exciton pairing at $\nu_1 = \nu_2 = 1/3$ re-emerges at $B = 12$~T. As shown in Fig.~\ref{fig3}c, all Jain states—including the Laughlin state at $\nu_1 = \nu_2 = 1/3$—display finite counterflow conductance (middle panel), accompanied by perfect drag responses (bottom panel). Compared to the high-field measurement at $B = 28$~T, the reduced magnetic field weakens the charge gap at $\nu_1 = \nu_2 = 1/3$, thereby allowing quasiparticle excitations at low temperature and facilitating exciton formation. A corresponding enhancement of the drag response at $\nu_1 = \nu_2 = 1/3$ is also observed in the Hall-bar device at $B = 12$~T, as illustrated in Fig.~\ref{EDL_hallbar}d.

Interestingly, the Laughlin state at $\nu_1 = \nu_2 = 1/3$ exhibits a charge gap of $\Delta = 12$~K at $B = 12$~T, which is substantially higher than the temperature of $\sim 20$~mK where the perfect drag response is observed. 
This observation strongly suggests that, while the energy cost of generating individual quasiparticle excitations from the Laughlin state is $\Delta = 12$~K, the pairing between positively and negatively charged constituents dramatically reduces the effective energy required to form excitons. 
Hence, $\Delta = 12$~K provides an additional estimate for the lower bound of the exciton binding energy at $B = 12$~T. Moreover, our findings suggest that exciton pairing between Laughlin states occurs by binding quasiparticle and quasihole excitations across graphene layers.

This hypothesis is validated by the temperature dependence of the drag response at $B = 28$~T and $\nu_1 = \nu_2 = 1/3$, where thermal activation induces a perfect drag response emerging from an insulating ground state. 
As shown in Fig.~\ref{fig3}d, the absence of excitons in the low-temperature regime results in vanishing $I_{\mathrm{drive}}$ and $I_{\mathrm{drag}}$, indicating that both graphene layers behave as insulators. 
With increasing temperature, both the drive and drag currents begin to rise around $T \approx 2$~K, which is far below the charge gap extracted from PF conductance (see Fig.~\ref{fig3}a). 
Remarkably, this reveals a narrow temperature window of perfect drag response, highlighted by the red-shaded region in Fig.~\ref{fig3}e. 
This observation provides compelling evidence for exciton pairing between thermally excited quasiparticles and quasiholes. 
Upon further increasing temperature, $I_{\mathrm{drive}}$ and $I_{\mathrm{drag}}$ bifurcate, and the drag ratio deviates from unity, indicating that exciton pairing is suppressed at elevated temperatures.

Notably, the interlayer excitons observed along the Jain sequence should be distinguished from those described by the $(111)$ state at $\nu_1 = \nu_2 = 1/2$~\cite{Eisenstein2014,Li2017superfluid,Liu2017superfluid,Liu2022crossover}, as well as from those associated with two-component fractional quantum Hall states~\cite{Zhang2025fractionalexciton}. 
Ground states characterized by two-component wavefunctions, such as the $(111)$ and $(333)$ states~\cite{Halperin1983,Wen1992}, emerge in regimes where strong interlayer correlations play a defining role in establishing the FQH phase. 
In contrast, the Laughlin state is stabilized predominantly by intralayer Coulomb interactions, while excitons in this regime arise from weak interlayer coupling between quasiparticle excitations.

\vspace{0.1 in}
\noindent\textbf{Exciton pairing at $3/8+3/8$ filling}

Having established exciton pairing in the Laughlin state, we now extend our investigation to other FQH states along the equal-density line defined by $\nu_1 = \nu_2 = \nu$. 
Fig.~\ref{fig4}a presents the parallel-flow conductance $G_{\mathrm{PF}}$ as a function of $\nu$ and $B$ along the equal-density line, revealing a transition between regimes of strong and weak interlayer correlation (see Fig.~\ref{EDL-B} for a full range map). 
FQH states with strong interlayer coupling appear at $B < 10$~T, where the positions of insulating states show excellent agreement with two-component FQH states with interlayer flux attachment~\cite{Li2019pairing,Zhang2025fractionalexciton}. 
As $B$ increases, a crossover from two-component to single-component FQH states leads to a suppression in the charge gap, producing a series of $B$-induced transitions at $\nu_1 = \nu_2 = 1/3$, $2/5$, and $3/7$, as indicated by the horizontal black lines in Fig.~\ref{fig4}a. 
In the high-field regime of weak interlayer correlation, the FQH states follow the characteristic hierarchical structure of the Jain series, consistent with wavefunctions that are single-component in nature.

Interestingly, in the regime of weak interlayer correlation, a FQH state emerges at $\nu_1 = \nu_2 = 3/8$, which lies outside the conventional Jain sequence~\cite{Pan.03,xia2004electron,pan2015fractional,bellani2010optical,samkharadze2015observation}. This state not only exhibits clear signatures of exciton pairing but also displays a drag response that is qualitatively distinct from that of the surrounding Jain fractions (Fig.~\ref{fig4}b--d). As illustrated in Fig.~\ref{fig4}c, the state at $\nu_1=\nu_2=3/8$ shows a substantially enhanced counterflow conductance compared with nearby fillings. Moreover, along the equal-density line, the $\nu_1=\nu_2=3/8$ state produces the most prominent counterflow drag signal (Fig.~\ref{fig4}e), underscoring the unusually strong interlayer correlations at this filling. Strikingly, the drag signal at $\nu_1=\nu_2=3/8$ is sharply confined to $\Delta\nu = 0$: as shown in Fig.~\ref{fig4}f, the drag response decays rapidly once $\Delta\nu$ departs from zero, in stark contrast to the broader response observed at $\nu_1 = \nu_2 = 1/3$ (Fig.~\ref{fig2}b--g).

Exciton pairing at $\nu_1 = \nu_2 = 3/8$ represents an unprecedented phenomenon. While the drag response by itself does not allow an unambiguous identification of the underlying ground state, we discuss below two plausible scenarios.

First, exciton pairing may be an intrinsic component of the underlying FQH state. Canonical examples include the $(111)$ state at $\nu_1=\nu_2=1/2$ and the $(333)$ state at $\nu_1=\nu_2=1/6$~\cite{Wen1992,Eisenstein2014,Zhang2025fractionalexciton}. Along this vein, the simplest construction for $\nu_1=\nu_2=3/8$ is an effective $(111)$-like state formed between half-filled $\Lambda$-levels (see Methods for further discussion). The precise structure of such a state, however, remains unclear and will require future theoretical work.

Alternatively, the exciton may share a similar construct to the Laughlin state at $\nu_1 = \nu_2 = 1/3$. 
In this scenario, the FQH state is primarily stabilized by intra-layer Coulomb interactions, while exciton pairing arises from binding a quasiparticle in one graphene layer to a quasihole in the other. 
Since this filling corresponds to a half-filled $\Lambda$-level of composite fermions~\cite{Pan.03,xia2004electron,pan2015fractional,bellani2010optical,samkharadze2015observation}, the FQH state is likely described by a non-Abelian wavefunction ~\cite{scarola2002possible,Mukherjee2012Pfaffian,mukherjee2014possible,Bonderson2008}. 
 
In the second scenario, quasiparticle excitations are non-abelian, yielding excitons that inherit non-Abelian statistics. The unusually strong interlayer correlations at this filling may therefore be a manifestation of non-Ableian topological order. Notably, although a FQH state at $3/8$ filling has recently been reported in ultra–high-mobility GaAs samples~\cite{wang2022even,wang2023next,wang2023fractional}, it has been conspicuously absent in monolayer graphene~\cite{Zibrov2018even,Zeng2019,Chen2019,Polshyn2018corbino}. As such, further theoretical and experimental investigations will be required to resolve the nature of this state.

Taken together, our findings reveal a new direction in which exciton pairing serves as the pathway for realizing novel quantum particles that behave like charge-neutral anyons. This new class of bilayer excitons opens a promising frontier for engineering exotic quantum phases, including anyonic superfluidity~\cite{Han2025anyon}, and is poised to stimulate broad future research efforts.

\section*{acknowledgments}
This material is based on the work supported by the Air Force Office of Scientific Research under award no. FA9550-23-1-0482. N.J.Z., R.Q.N., and J.I.A.L. acknowledge support from the Air Force Office of Scientific Research. N.J.Z. acknowledges partial support from the Jun-Qi fellowship. R.Q.N. and J.I.A.L. acknowledge partial support from the National Science Foundation EPSCoR Program under NSF Award OIA-2327206. Any opinions, findings and conclusions or recommendations expressed in this material are those of the author(s) and do not necessarily reflect those of the National Science Foundation. This work was performed in part at the Aspen Center for Physics, which is supported by a grant from the Alfred P. Sloan Foundation (G-2024-22395). A portion of this work was performed at the National High Magnetic Field Laboratory, which is supported by National Science Foundation Cooperative Agreement No. DMR-1157490 and the State of Florida. N.B. and D.E.F. were supported in part by the National Science Foundation under Grant No. DMR-2204635. S.A. acknowledges support from Kuwait University. K.W. and T.T. acknowledge support from the JSPS KAKENHI (Grant Numbers 21H05233 and 23H02052) and World Premier International Research Center Initiative (WPI), MEXT, Japan. Part of this work was enabled by the use of pyscan (github.com/sandialabs/pyscan), scientific measurement software made available by the Center for Integrated Nanotechnologies, an Office of Science User Facility operated for the U.S. Department of Energy.

\section*{Data availability}
The data that support the plots within this paper and other findings of this study are available from the corresponding author upon reasonable request.

\section*{Author contribution}
N.J.Z. and J.I.A.L. conceived the project. N.J.Z., R.Q.N., and
X.L. fabricated the device. N.J.Z., R.Q.N., S.A. performed the measurement. N.B. and D.E.F. provided theoretical inputs. K.W. and T.T. provided the material. N.J.Z., R.Q.N., N.B., S.A., D.E.F., and J.I.A.L wrote the manuscript together.

\section*{Competing financial interests}
The authors declare no competing financial interests.

%\bibliography{Li_ref}%
\bibliography{Li_ref}

\newpage
\clearpage

\section*{Method}

\renewcommand{\thefigure}{M\arabic{figure}}
\def\theequation{M\arabic{equation}}
\def\thetable{M\Roman{table}}
\setcounter{figure}{0}
\setcounter{equation}{0}

In this section, we provide detailed discussions to further substantiate results reported in the main text. This section offers a comprehensive review, summarizing the notations employed and elaborating on the measurement configurations as well as the inter-layer degrees of freedom involved in the experiments. The theoretical methods section explores the relevant wave functions at integer and fractional fillings of the $\Lambda$-levels, the nature of quasiparticles, and the quantum statistics of excitons arising from pairing between quasiparticles and quasiholes. For additional information and in-depth analysis, readers are directed to the Supplementary Materials ~\cite{SI}.

This work employs a suite of transport measurements made possible by patterning quantum Hall bilayer graphene into distinct device geometries. In the sections below, we describe the details of these measurement techniques.

\subsection{I. Edgeless Double Corbino Geometry}

The edgeless Corbino geometry enables bulk transport measurements without the influence of edge conduction. We utilize three measurement configurations---parallel flow (PF), counterflow, and drag ~\cite{Li2019pairing,Zeng2023solid,Zhang2025fractionalexciton}. In PF and counterflow, equal currents are applied to both graphene layers, flowing in the same or opposite directions, respectively. In the drag configuration, a current bias is applied to the top layer, which induces a current response in the bottom layer.

These complementary transport configurations allow us to identify interlayer excitons. In the PF geometry, a FQHE state manifests as an insulating response with vanishing bulk conductance, $G_{\mathrm{PF}} = 0$, reflecting the absence of free charge carriers due to the robust charge gap in both graphene layers. The simultaneous presence of a bilayer exciton condensate and a charge gap leads to a finite counterflow conductance, $G_{\mathrm{counter}} > 0$, even while $G_{\mathrm{PF}}$ remains zero. This finite counterflow response arises from an exciton flow that generates equal and opposite electrical currents in the two layers. Because excitons are overall charge neutral, such a flow is compatible with a charge-gapped state.

The drag configuration provides an additional and unambiguous signature of exciton transport through the observation of a ``perfect drag'' response ~\cite{Nandi2012exciton,Eisenstein2014}. Here, the current in the drive layer is mirrored with equal magnitude and opposite sign in the drag layer, such that
\[
\frac{I_{\mathrm{drag}}}{I_{\mathrm{drive}}} = 1.
\]

In this work, the corbino-shaped quantum Hall bilayer sample consists of a $4.5$\,nm thick hexagonal boron nitride (hBN) separation layer.

\begin{figure*}
\includegraphics[width=0.85 \linewidth]{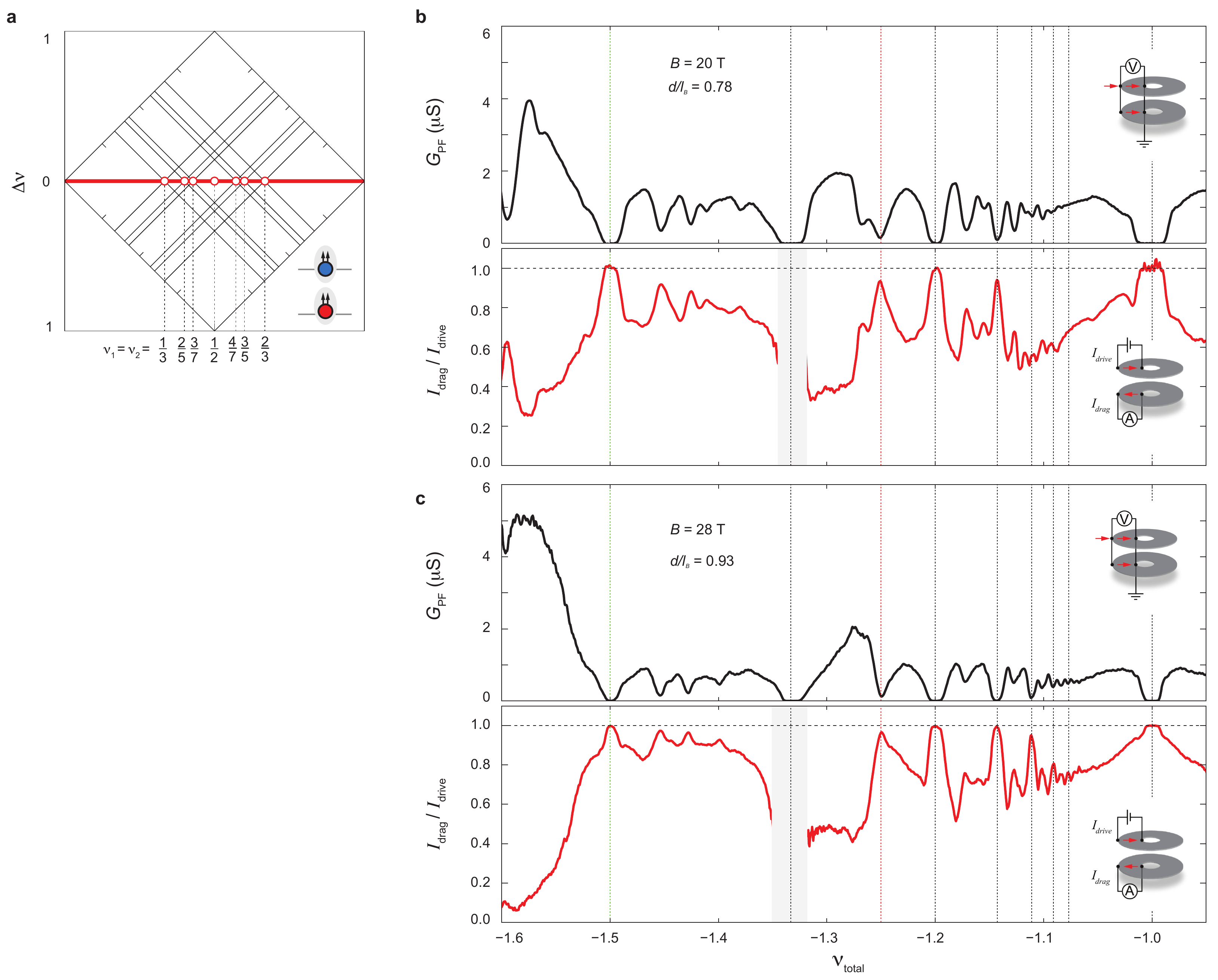}
\caption{\label{EDL_corbino}
{\bf Excitonic pairing in the Corbino geometry.} 
(a) Schematic phase diagram of a quantum Hall bilayer, parameterized by $\nu_{\mathrm{total}}$ and $\Delta\nu$. 
Black solid lines mark the expected trajectories of layer-decoupled FQH states belonging to the Jain sequence. 
The red solid line denotes the equal-density line, where the filling factors in the two graphene layers are equal, $\nu_1 = \nu_2$. 
Red circles indicate the expected locations of Jain states along this line. 
(b--c) Parallel-flow conductance $G_{\mathrm{PF}}$ (top) and drag ratio $I_{\mathrm{drag}}/I_{\mathrm{drive}}$ (bottom) measured along the equal-density line as a function of $\nu_{\mathrm{total}}$ at (b) $B = 20$~T and (c) $B = 28$~T. 
The emergence of FQH states is signaled by vanishing $G_{\mathrm{PF}}$, whereas exciton pairing is manifested by perfect drag response with $I_{\mathrm{drag}}/I_{\mathrm{drive}} = 1$. 
Black vertical dotted lines mark the expected positions of Jain-sequence FQH states, while the red vertical dotted line denotes a half-filled $\Lambda$-level. 
The green dotted line marks the location of the two-component FQH state identified in previous works~\cite{Li2019pairing,Liu2019interlayer,Zhang2025fractionalexciton}.
}
\end{figure*}

Fig.~\ref{EDL_corbino} shows the parallel-flow and drag responses measured along the equal-density line in the Corbino-shaped sample. 
Perfect drag response is observed across a series of FQH states, each accompanied by a vanishing $G_{\mathrm{PF}}$.

Notably, the Corbino-shaped sample exhibits exceptionally strong interlayer correlations, evidenced by an abundance of FQH states displaying perfect drag response, persisting up to $d/\ell_B = 0.93$. 
This behavior stands in stark contrast to that of the Hall-bar device, as discussed below.

%The full heterostructure is encapsulated by hBN and dual graphite gates. This dual-gated design allows independent tuning of the charge carrier densities in layers 1 and 2, $n_{1}$ and $n_{2}$, via voltages applied to the respective graphite gates. The Landau level (LL) filling factor in each layer is given by $\nu_i = n_i \Phi_{0}/B$, where $\Phi_{0}$ is the magnetic flux quantum and $B$ is the external magnetic field.

\subsection{II. Counterflow Drag in the Hall-Bar Geometry}

\begin{figure*}
\includegraphics[width=0.7\linewidth]{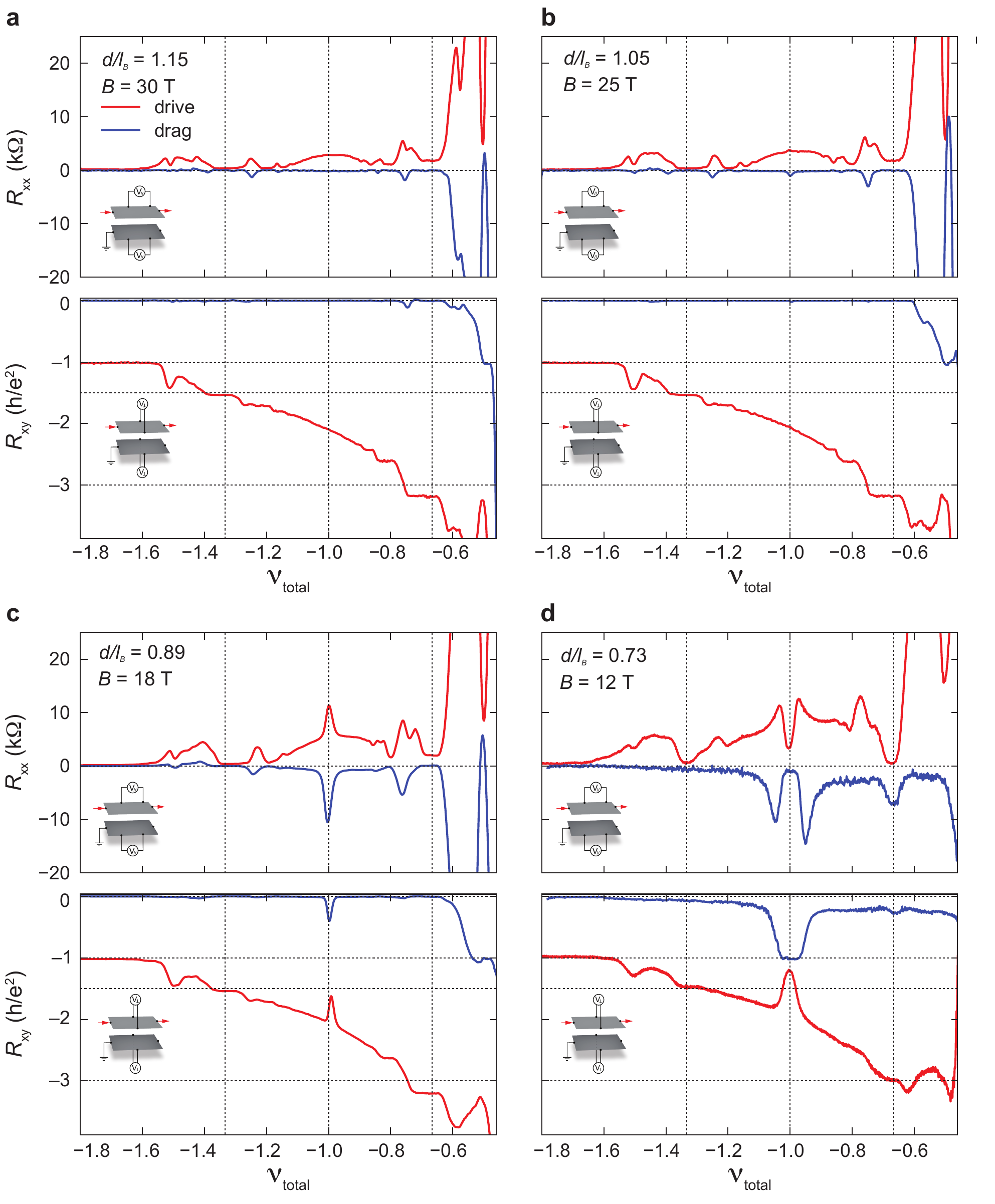}
\caption{\label{EDL_hallbar}
{\bf Counterflow drag measurements in the Hall-bar–shaped sample.} 
Longitudinal and Hall resistances, $R_{xx}$ (top) and $R_{xy}$ (bottom), of the drive (red) and drag (blue) layers measured along the equal-density line as a function of $\nu_{\mathrm{total}}$ at 
(a) $B = 30$~T, 
(b) $B = 25$~T, 
(c) $B = 18$~T, and 
(d) $B = 12$~T. 
Black vertical dashed lines indicate the expected locations of the Laughlin states at $\nu_{\mathrm{total}} = -4/3$ and $-2/3$, as well as the (111) state at $\nu_{\mathrm{total}} = -1$. An enhancement of drag response at $\nu_1=\nu_2=1/3$  is observed at $B = 12$ T (panel d).
}
\end{figure*}

A second device, patterned into a Hall-bar geometry, is also used in this work. This sample incorporates a $5.4$\,nm thick hBN layer as the interlayer insulating barrier. In Hall-bar quantum Hall bilayers, signatures of exciton pairing and interlayer correlations are well established through counterflow drag measurements~\cite{Li2017superfluid,Liu2017superfluid,Li2019pairing,Liu2019interlayer}. For instance, an exciton condensate at $\nu_1=\nu_2=1/2$ is identified by a quantized Hall response in which both the drive and drag layers exhibit the same Hall plateau, as illustrated in Fig.~\ref{EDL_hallbar}d.

Fig.~\ref{EDL_hallbar} shows counterflow drag measurements performed along the equal-density line at various values of $d/\ell_B$ in the Hall-bar–shaped device. 
Here, the quantized Hall plateaus in the drive layer provide an unambiguous identification of the Jain-sequence FQH states.

\begin{figure*}
\includegraphics[width=0.5\linewidth]{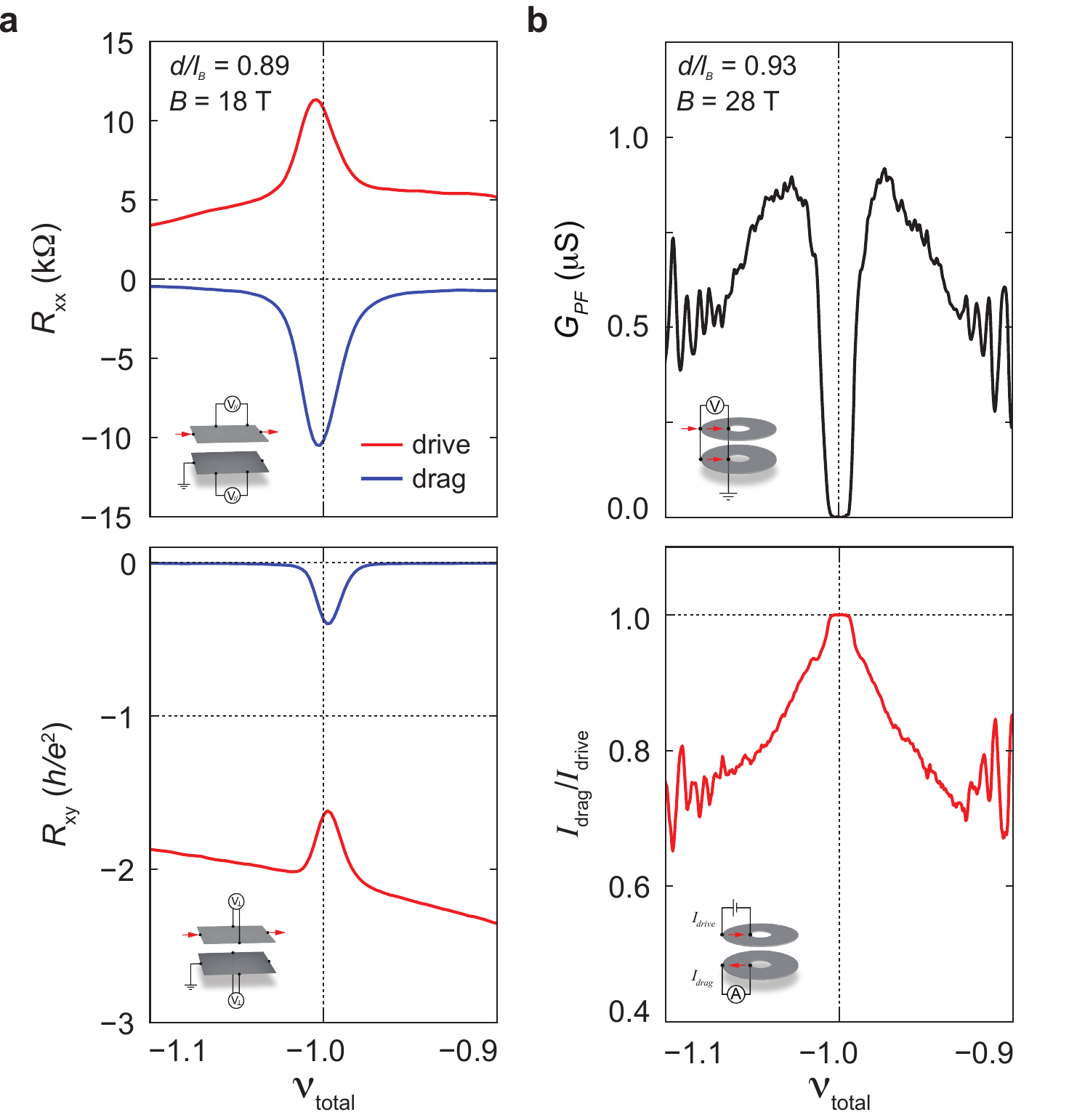}
\caption{\label{fig:hallbar_corbino_comparison} \textbf{Comparison of Hall bar and Corbino geometries.}
(a) Longitudinal (top) and Hall (bottom) resistances $R_{xx}$ and $R_{xy}$ of a double-layer device in a Hall bar geometry measured as a function of $\nu_\mathrm{total}$ at $B = 18$~T for drive (red) and drag (blue) layers. (b) Parallel-flow conductance $G_\mathrm{PF}$ (top) and drag ratio $I_\mathrm{drag}/I_\mathrm{drive}$ (bottom) of a double-layer device in a Corbino geometry measured as a function of $\nu_\mathrm{total}$ at $B = 28$~T. }
\end{figure*}

\begin{figure*}
\includegraphics[width=0.88\linewidth]{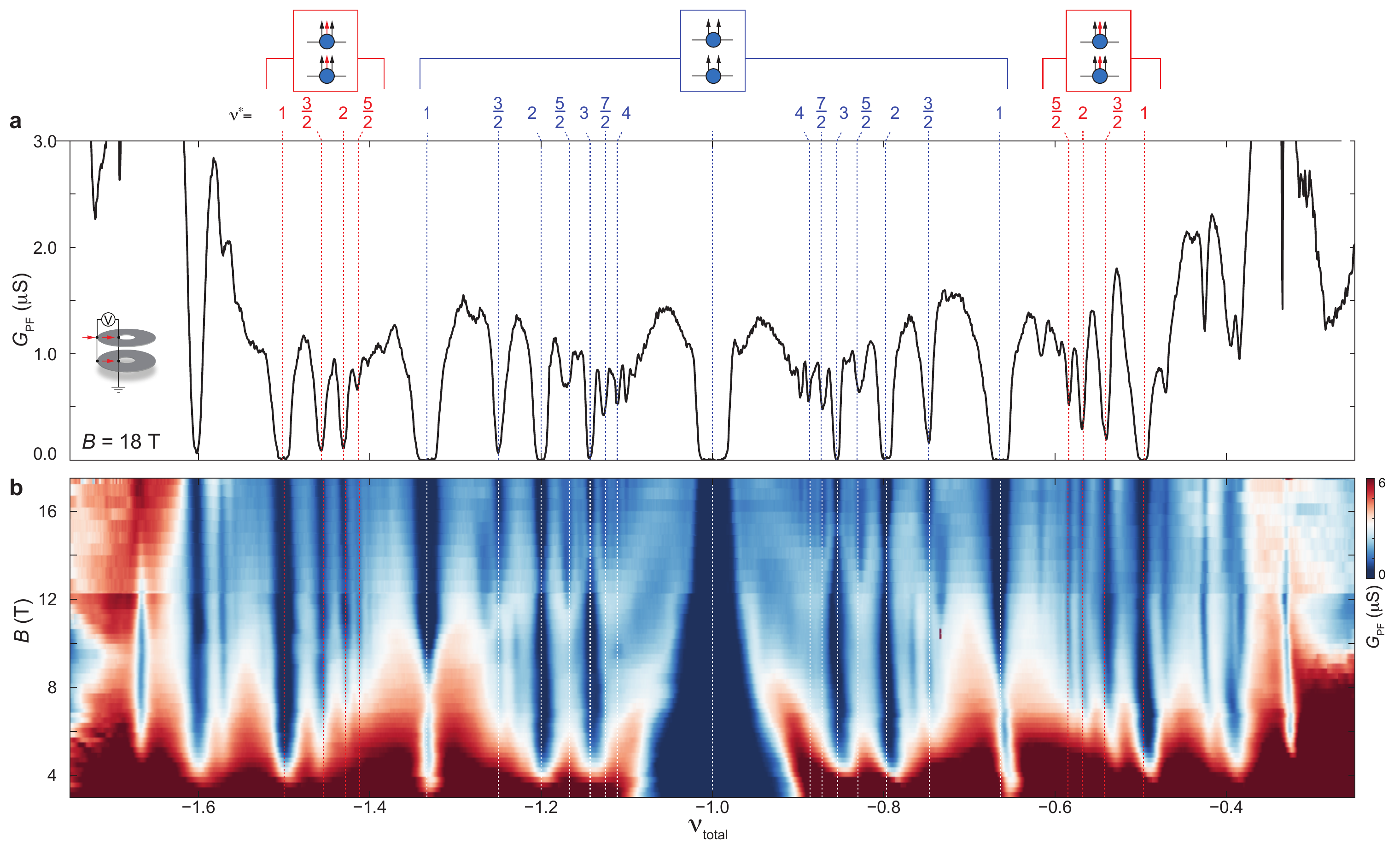}
\caption{\label{EDL-B}{\bf Field-induced transition between two-component and one-component FQH states.}
(a,b) Parallel-flow conductance $G_\mathrm{PF}$ measured along the equal-density line of the Corbino-shaped device, shown (a) as a function of $\nu_{\mathrm{total}}$ at $B = 18$~T and (b) as a color scale map with varying $\nu_{\mathrm{total}}$ and $B$. FQH states occurring at integer and half-integer composite-fermion fillings are indicated by vertical dotted lines. Within the range $-4/3 < \nu_{\mathrm{total}} < 2/3$, the observed FQH states follow the sequence with one interlayer flux attachment at low field ~\cite{Li2019pairing}, before transitioning to the Jain sequence without interlayer flux attachment at higher field ~\cite{Jain.03}. For $\nu_{\mathrm{total}} < -4/3$ and $\nu_{\mathrm{total}} > 2/3$, the sequence with one interlayer flux attachment persists up to $B = 18$~T ~\cite{Li2019pairing}.}
\end{figure*}

Fig.~\ref{fig:hallbar_corbino_comparison} highlights the transport response at $\nu_{\mathrm{total}} = -1$, where an exciton condensate is expected, measured using the two different device geometries. 
Both the Hall-bar and Corbino devices are tuned to an effective interlayer separation of $d/\ell_B = 0.9$, a regime in which similar interlayer correlations are anticipated. Interestingly, the drag response is much weaker in the Hall-bar–shaped sample. 
While the Corbino-shaped sample exhibits perfect drag response together with a fully developed charge gap (vanishing $G_{\mathrm{PF}}$), the Hall-bar sample shows only partial signatures of exciton pairing—manifested as peaks in the longitudinal and transverse drag signals—indicating that the exciton condensate is not fully developed. 
This discrepancy is consistent with previous reports, which found that bulk transport in the Corbino geometry is more sensitive to fragile electronic states~\cite{Zeng2019,Li2019pairing}.

%Notably, tuning $\nu_{\text{total}}$ at \dnu $=0$ results in a region with a zero drag ratio, outlined by blue dashed lines and highlighted by the blue-shaded area in the middle and right panels of Fig.~\ref{fig3}b, respectively. Fig.~\ref{fig3}e plots the transport responses as $\nu_1$ is detuned from the Laughlin state. The onset of $G_{\text{PF}}$ indicates that the charge gap is suppressed by the presence of quasiparticles in layer 1 (black solid line in Fig.~\ref{fig3}e). However, the absence of quasiholes in layer 2 prevents the formation of exciton pairs, thereby suppressing the counterflow conductance, $G_{\text{counter}}$ (blue solid line in Fig.~\ref{fig3}e). This results in a non-zero $G_{\text{PF}}$ and a zero $G_{\text{counter}}$, confirming that exciton formation requires doping with varying $\Delta \nu$ while maintaining a constant $\nu_{\text{total}}$.

\subsection{III. Theoretical Methods}

The structure of a fractional quantum Hall wavefunction in the lowest Landau level is highly constrained on the grounds of analyticity requirements. By representing the {electron coordinates} as $z_a=x_a+iy_a$, the wavefunction {takes} the form,
{\begin{align}
\psi(z_1,z_2,\dots) = \Phi(z_1,z_2\dots) \exp\left( - \sum_{n}\frac{|z_n|^2}{4l_B^2} \right),
\end{align}}  
where $l_B$ is the magnetic length, and $\Phi(z_1,z_2,\dots)$ is an antisymmetric holomorphic function.  Note that a more realistic wavefunction may not have the above form, for example, due to LL mixing. However, for the purpose of discussing the topological properties of the ground state, we use the above description as a representative form of the true wavefunction.  In a bilayer system, electron coordinates may be distinguished based on their layer index, thus we include an additional index, {$z_{i,n}=x_{i,n}+iy_{i,n}$} with $i=1,2$ representing the two layers. Since the exponential is a product in all coordinates, the holomorphic function $\Phi(z_1,z_2,\dots)$ encodes all the information about the electron correlations of the FQHE state. The nature of the ground state in any actual sample depends on an intimidating list of details including Coulomb interactions, screening, and disorder that makes the problem of determining $\Phi$ extremely challenging. Nevertheless, when solely focusing on the topological properties of the ground state, one is able to write down representative wavefunctions which fall in the same universality class as the true wavefunctions~\cite{Wen2004quantum}. 

All the allowed wavefunctions for the FQHE states can be divided into two groups: Abelian and non-Abelian wavefunctions. As is relevant to the data, we first focus on the bilayer Abelian states described by the integer filled ${}^2_0$CF levels. Then, the structure of the bilayer Abelian wavefunctions is encoded in a $2\times 2$ block matrix ${\bf K}$, consisting of blocks {$[{\bf K}_{ij}]$} that describe intra-layer flux attachments when $i=j$, and interlayer flux attachments when $i\ne j$. The diagonal blocks {$[{\bf K}_{ii}]$} are then simply the monolayer symmetric integer ${\bf K}$ matrices for each layer. {Their dimensions $\dim{[{\bf K}_{ii}]}$ represent the number of different pseudo-spin indices for each layer that can be labeled by a superscript $\sigma_i=1,\dots,\dim \left[{\bf K}_{ii}\right]$}. The holomorphic function $\Phi$ is constructed as,
\textcolor{black}{\begin{align}
\Phi(\{z^{\sigma_1}_{1,n}\},\{z^{\sigma_2}_{2,m}\}) = \prod_{i,j,\sigma_i,\sigma_j} \prod_{n,m} (z^{\sigma_i}_{i,n}-z^{\sigma_j}_{j,m})^{\frac{[{\bf K}_{ij}]_{\sigma_i\sigma_j}}{2}},
\end{align}}
which has a natural construction in terms of the bilayer ${}^a_b$CFs \cite{Zhang2025fractionalexciton}. In the case of ${}^2_0$CFs, the bilayer ${\bf K}$-matrix becomes block diagonal and the bilayer wavefunction becomes a product of two monolayer FQHE states. The non-Abelian states, on the other hand, require a more careful construction involving conformal field theory. In the following, we focus on the observed ${}^2_0$CF bilayer states and discuss their likely exciton compositions.

\subsubsection{Integer filled ${}^2_0$CF level states}
 
In this section, we focus our attention on the integer filled ${}^2_0$CF level states, i.e., $\nu_i^*\in \mathbb{Z}$ for $i=1,2$ representing the layer index. The ${}^2_0$CF levels $\nu_i^*$ relates to the true filling fractions $\nu_i$ of the $i$th layer as $\nu^*_i=\nu_i/(1-2\nu_i)$. The absence of interlayer flux permits a block diagonal representation of the bilayer ${\bf K}$-matrices as ${\bf K} = \text{diag}({\bf K}_1,{\bf K}_2)$, where ${\bf K}_i$ is the monolayer ${\bf K}$-matrix corresponding to the $i$th layer, along with a charge vector ${\bf t}_i$. Working in the symmetric representation, such that the charge vectors ${\bf t}^T_i=(1,\dots,1)$, the monolayer blocks take the form ${\bf K}_i = \sigma_i {\bf I} + 2 {\bf J}$ with $\sigma_i=\pm 1$, and the matrices ${\bf I}$ and ${\bf J}$ are the identity and a constant matrix of ones \cite{kane1995impurity}. Notice that $\sigma_i$ is chosen depending on the ${}^2_0$CF particle-/hole-doped $\Lambda$ levels so we identify $\sigma_i=\text{sign}(\nu_i^*)$. The dimension of each of the blocks is set by the charge vector $\dim{\bf t}_i$ which represents the number of $U(1)$ gauge fields in the bulk Chern-Simons field theoretic description. The block diagonal structure allows straightforward manipulation, such as the filling fractions of each layer is given by $\nu_i = {\bf t}_i^{T}{\bf K}_i^{-1}{\bf t}_i$ for $i=1,2$. The total and the difference of fillings are defined as $\nu_{\text{total}}=\nu_1+\nu_2$ and $\Delta\nu=\nu_1-\nu_2$ respectively. 

Having a well-defined ${\bf K}$-matrix description of the system allows one to evaluate the quasiparticle content of the topological order \cite{Wen2004quantum}. The structure of the ${\bf K}$-matrix determines the coupling structure of the $U(1)$ gauge fields in the low-energy Chern-Simons description. Localized quasiparticle excitations are then created by inserting a flux quantum for each gauge field. Since there are a total of $\dim {\bf t}_1+\dim {\bf t}_2$ gauge fields, we label the flux insertion by an integer vector ${\bf l}$ of dimension $\dim {\bf l}=\dim ({\bf t}_1\oplus {\bf t}_2)$. The minimal quasiparticle charge corresponding to ${\bf l}$ defines the fundamental quasiparticle excitation of the system. The quasiparticle charge in each layer is given by $Q_1=-e({\bf t}_1\oplus {\bf 0})^T{\bf K}^{-1}{\bf l}$ and $Q_2=-e({\bf 0}\oplus {\bf t}_2)^T{\bf K}^{-1}{\bf l}$.
In general since a quasiparticle excitation is labeled by the vector ${\bf l}$, a generic excitation of the system may have its charge localized in both of the layers \cite{Zhang2025fractionalexciton}. However, due to the block diagonal structure of the bilayer ${\bf K}$-matrix, the quasiparticle charge in each layer decouples and can be independently created by flux insertions labeled by integer vectors ${\bf l}_i$ of dimensions $\dim{\bf l}_i=\dim {\bf t}_i$, such that ${\bf l}={\bf l}_1\oplus{\bf l}_2$. This is a direct consequence of absence of interlayer flux in the bilayer ${}^2_0$CF construction. 

The localized quasiparticle charge, for a bilayer ${}^2_0$CF state, in each layer can then be written as,
\begin{align}
Q_i = -e~ {\bf t}_i{\bf K}^{-1}_i{\bf l}_i,
\end{align}
for $i=1,2$. The fundamental charge of each layer is then $Q_{\text{min.},i}=\pm e/|2\nu_i^*+1|$ (see Supplementary Materials for a derivation). Note that we used the symmetric representation of monolayer ${\bf K}$-matrices which admit a simple form of inverse given by,
\begin{align}
{\bf K}_i^{-1} = \sigma_i {\bf I} - \frac{2\sigma_i}{2\dim {\bf K}_i + \sigma_i} {\bf J},
\end{align} 
where again the matrices ${\bf I}$ and ${\bf J}$ are the identity and a constant matrix of ones. Exciton charge is computed from the condition $Q_1+Q_2=0$ for a minimal $Q_{1,2}\ne 0$. 
The resulting exciton {\color{black}dipole} charges are found to be,
\begin{align}
Q_{\text{exciton}} =\pm \frac{ e}{\text{gcd}(|2\nu_1^*+1|,|2\nu_2^*+1|)},
\end{align}
where the integer valued function $\text{gcd}(n_1,n_2)$ computes the greatest common divisor of the integers $n_1$ and $n_2$. 

Similar to the quasiparticle charge, the bilayer ${\bf K}$-matrix also encodes the exchange statistics for its quasiparticle content. The self statistical phase of a quasiparticle excitation, labeled by ${\bf l}$, upon an exchange with itself is given by $\theta_s = \pi {\bf l}{\bf K}^{-1}{\bf l}\mod 2\pi$  \cite{Wen2004quantum,Zhang2025fractionalexciton}. For the special case of ${}^2_0$CF family, when the bilayer ${\bf K}$-matrix is block diagonal, we obtain $\theta_s = \pi \sum_{i=1}^{2}{\bf l}_i{\bf K}^{-1}_i{\bf l}_i\mod 2\pi$, where ${\bf l}={\bf l}_1\oplus{\bf l}_2$. Notice that for a given quasiparticle charge $Q$, the equation $Q=-e\sum_{i=1}^2 {\bf t}^T_i{\bf K}_i^{-1}{\bf l}_i$ may have distinct solutions of ${\bf l}$. However, all such quasiparticles, corresponding to the distinct ${\bf l}$ but with same charge, acquires the same statistical phase (modulo $2\pi$) upon exchange (For a proof, see supplementary Materials of Ref. \cite{Zhang2025fractionalexciton}). Therefore, the self-statistical phase of ${}^2_0$CF bilayer layer excitons at filling $\nu_1^*$ and $\nu_2^*$ (see Supplementary Information) is given by, 
\begin{align}
\frac{\theta_s}{\pi} =  \sum_{i=1}^2 \frac{\left(|2\nu_i^*+1|-2\right)\left(2\nu_i^*+1\right) }{\text{gcd}(|2\nu_1^*+1|,|2\nu_2^*+1|)^2}\mod 2.
\end{align}
{In general, the charge neutral ${}^2_0$CF interlayer excitons follow non-trivial (Abelian) anyonic statistics, however, at certain fillings, for example when $|2\nu^*_1+1|$ and $|2\nu^*_2+1|$ are co-prime, the statistics is trivial. Indeed one can show that when such excitons are trivial, they follow Bosonic statistics (see Supplementary Information).}
%Clearly, the charge neutral ${}^2_0$CF interlayer excitons follow non-trivial (Abelian) anyonic statistics. 
Note that the above statistical phase is the phase acquired on a single exchange of two excitons {\color{black} of the same kind}. This contrasts with the braiding phase which is the phase acquired on a single counter-clockwise winding of one exciton around another {\color{black} of the same kind}. Topologically, this is achieved by two consecetive exchanges thus the braiding phase is twice the exchange phase. The present analysis assumes integer effective filling, i.e., $\nu_i^*\in\mathbb{Z}$, however, some of the observed states also occur at fractionally filled effective CF filling which we discuss in the next section.

\subsubsection{{Bilayer $3/8+3/8$ state}}
%\subsubsection{Fractionally filled ${}^2_0$CF level states}

%Along with the integer filled ${}^2_0$CF level states, we also observe a state corresponding to the half filled $\Lambda$ level of the ${}^2_0$CF construction, at $\nu_i^*=3/2$ for $i=1,2$. 

%{\color{red} At any filling factor, an infinite number of theoretically possible states can be constructed. Physically relevant states tend to have a simple physical interpretation in terms of composite fermions. For our filling factor, one approach to construct a state would generalize the $(111)$ construction at the filling factor of 1. At the same time, it is possible that the unusual behavior at the total filling factor $3/4$ reflects non-Abelian statistics. Below, we discuss two possible $3/8+3/8$ states. One generalizes the $(111)$ state and the other is non-Abelian.}

Just as with any other filling factor, an infinite number of theoretically possible states can, in principle, be constructed at $\nu_1=\nu_2=3/8$. Here, we highlight two representative scenarios. In the first, the system resembles an effective (111) exciton condensate at total filling $\nu_{\mathrm{tot}}=1$, generalized to the fractional regime. In the second, each layer independently forms an even-denominator state, and exciton pairing is subsequently constructed between the resulting quasiparticles and quasiholes across the two layers. We emphasize that these examples are not intended to exhaust all possible wavefunctions at this filling, but rather to provide context for the novelty of exciton pairing observed at this unique fraction.

Along with the integer filled ${}^2_0$CF level states, \textcolor{black}{the data show drag response at fractions} corresponding to the half-filled $\Lambda$ level of the ${}^2_0$CF construction. \textcolor{black}{More precisely these occur at $\nu_i=3/8,5/8,5/12$, and $7/12$ that translate to an integer number of fully filled, along with a single half filled, $\Lambda$ levels in each layer ($i=1,2$). Some of these states have also been reported in ultra-high mobility GaAs samples, presumably originating from the residual composite Fermion interaction\cite{wang2022even,wang2023fractional,wang2023next}. Notably, these states have not yet been observed in monolayer Graphene samples with similar doubly-encapsulated geometry \cite{Zibrov2018even,Zeng2019,Chen2019,Polshyn2018corbino}. In this section, we provide different possible interpretations of these exotic states, some of which are non-Abelian in nature. For concreteness, we focus on the bilayer state at $\nu=3/8$ in each layer. In the ${}^{2}_{0}$CF construction, this state is obtained by filling $\nu^*=3/2$ $\Lambda$ levels. The entire discussion presented in this section can easily be extended to other even denominator states observed in the data by including more integer filled (and their particle/hole conjugate) $\Lambda$ levels.} 

\textcolor{black}{The parallel flow and drag measurement response of these exotic states provide some hints to their underlying nature. Since the bilayer system has a charge mobility gap, this bilayer state consisting of two half-filled $\Lambda$ levels can be viewed as a generalization of the Halperin (111) state. Indeed the combination of the two half-filled $\Lambda$ levels leads to a fully \textcolor{black}{charge-}gapped system. In terms of composite fermions, this interpretation forces us to view the dressed particles in the fully-filled $\Lambda$ levels differently from the ones occupying the half-filled $\Lambda$ level in each layer. With this in mind, we can pair the two half-filled $\Lambda$ levels and write down a bilayer ${\bf K}_{\text{CF}}$ (and the physical ${\bf K}={\bf K}_{\text{CF}} + 2{\bf J}_{2} \oplus {\bf J}_2$) matrix,
\begin{align}\label{eq:Kmat_111generalization}
{\bf K}_{\text{CF}} &=  \begin{pmatrix}
        1 & 0 & 0 & 0 \\
        0 & 1 & 1 & 0 \\
        0 & 1 & 1 & 0 \\
        0 & 0 & 0 & 1
    \end{pmatrix} ~~ \Rightarrow~~ {\bf K} & = \begin{pmatrix}
        3 & 2 & 0 & 0\\
        2 & 3 & 1 & 0\\
        0 & 1 & 3 & 2\\
        0 & 0 & 2 & 3
    \end{pmatrix},
\end{align}
along with the charge vector ${\bf t}^T=(1~1~1~1)$. One finds that the total filling $\nu_t$ and the difference of the fillings $\Delta \nu$ for this state works out as $\nu_t=3/4$ and $\Delta \nu=0$. One may also view each of the four levels as a separate layer and couple a fictitious external electromagnetic field to each of them and obtain its charge response. This results in a definition for the filling fraction for each level which works out as $\nu=1/4$ for the single fully filled $\Lambda$ level in each layer, and $\nu=1/8$ for the half filled $\Lambda$ level in each layer. Together this gives $\nu=1/8+1/4=3/8$ per layer as required.}

\textcolor{black}{In the Halperin (111) state, the perfect drag response is attributed to the exciton-condensate  {\color{black}ground state as opposed to excitations}\cite{Zhang2025fractionalexciton}. We now check whether the above state also has the same property. Indeed, the charge current contribution due to the half-filled $\Lambda$ level in the drive layer will induce an effective electric field in the drag layer due to the presence of interlayer fluxes between the two half-filled levels. One may then generalize the analysis provided previously (see Methods in \cite{Zhang2025fractionalexciton}) to systems with multiple $\Lambda$ levels with a general structure of flux attachment between them. \textcolor{black}{{\color{black} Note that one} cannot  assume the conductances of the two $\Lambda$ levels in each layer to be the same. Since the bilayer system is layer-symmetric, we call the distinct longitudinal conductances $\sigma_{xx}^{\text{I}}$ for the fully filled and $\sigma_{xx}^{\text{II}}$ {\color{black}for} the half-filled $\Lambda$ level. {\color{black} We expect that $\sigma_{xx}^{II}\gg\sigma_{xx}^I$.} One may view the state given in Eq. (\ref{eq:Kmat_111generalization}) as consisting of four different levels each with an effective {\color{black}composite-fermion} filling of $\nu_i=1$ ($i=1,\dots,4$) as a result of the general flux attachment. Then one obtains a relation between $j_{\text{drive}}$ and $j_{\text{drag}}$ currents (see Supplementary Information),
\begin{align}
j_{\text{drag}} = \frac{4\sigma^{\text{I}}_{xx}-15\sigma^{\text{II}}_{xx}}{4\sigma^{\text{I}}_{xx}+17\sigma^{\text{II}}_{xx}} j_{\text{drive}}.
\end{align}
Interestingly, the drag ratio $j_{\text{drag}}/j_{\text{drive}}\ne 1$ for any value of the conductances, therefore, the backflow current solely due to the condensate does not explain the perfect drag response. %We, %therefore, ascribe the perfect drag response to excitons. 
{\color{black} At the same time, at $\sigma_{xx}^I=0$, the ratio is $-15/17\approx -1$.}
}}
%The charged excitations in this state can be found by including source terms to the bulk theory. Each localized source of charge $Q_1$ , labeled by an integer vector $\textbf{l}$, is then given $Q=-e\left( {\bf t}_1 \oplus {\bf t}_2 \right)^{T} {\bf K}^{-1} {\bf l}$. We are interested in solutions for $Q$}

\textcolor{black}{We now discuss another possibility that leads to a non-Abelian structure. It should be noted that LL mixing plays a significant role in facilitating interaction between composite fermions, creating favorable conditions for pairing in half-filled $\Lambda$ levels \cite{wang2023next,Zibrov2018even}. Indeed, these unconventional states in GaAs were observed in hole-doped LL filling rather than electron-doped, presumably because the larger effective mass of holes, and consequently larger LL mixing, enhances interactions between CFs \cite{wang2022even,wang2023next,wang2023fractional}. In the present case of quantum Hall graphene bilayers, one may expect that the unique Coulomb environment reduces short-range repulsion in the CF sea, making it more favorable for CFs to form Cooper pairs. At large magnetic field, the absence of interlayer correlations restricts Cooper pairing to occur within each individual layer. As a result, the bilayer system offers a favorable platform to stabilize novel pairing states of composite fermion driven by intra-layer correlations, which are distinct from those observed in monolayer graphene samples.}

\textcolor{black}{In the present case of $\nu_i=3/8$ state in the lowest Landau level for each of the graphene layers,} previous numerical studies suggest that the ground state wavefunction of the monolayer $\nu=3/8$ in the lowest Landau level is described by the wavefunction with anti-Pfaffian pairing of four-flux composite fermions in the excited half-filled $\Lambda$ level \cite{scarola2002possible,Mukherjee2012Pfaffian,mukherjee2014possible,Bonderson2008}. The wavefunction belonging to the anti-Pfaffian topological class has been shown to have non-Abelian structure, similar the to one of $\nu=5/2$ candidate states. In this section, we give a brief summary of the structure of the wavefunction, \textcolor{black}{as is recognized for monolayer platforms in Ref. \cite{hutasoit2017enigma},} and the \textcolor{black}{expected excitonic content present in the bilayer $\nu_1=\nu_2=3/8$ state}, along with their topological properties. To begin, we first notice that since there are no interlayer fluxes, the excitons are formed by pairing a quasiparticle {\color{black}and a} quasihole in the two layers. We, therefore, first focus on the quasiparticle content of monolayer $\nu=3/8$ state. 

A topological field theory that describes the low energy physics of the FQHE states develops a gauge anomaly at the edge, which can be made gauge invariant by including dynamical degrees of freedom living on the edge of the FQHE liquid. This dynamical gauge theory is a chiral conformal field theory (CFT), which may also be used to construct candidate bulk wavefunctions of the corresponding FQHE state. Constructing bulk wavefunctions using this approach relies on the conjectured bulk-edge correspondence, which may break down for more complicated states \cite{hansson2017quantum}. Assuming the correspondence holds, the structure of the non-Abelian wavefunctions can then be studied by focusing on the edge. For the Abelian case, the ${\bf K}$-matrix simply describes the independent $U(1)$ bosonic modes living on the edge, and the quasiparticles are vertex operators in the bosonic sector. To describe the non-Abelian states, in addition to the charged sector comprising the bosonic modes, a neutral sector constructed from a chiral (or anti-chiral) part of the CFT is also included. Consequently, the quasiparticles are described by the products of vertex operators from the bosonic sector and the primary operators of the CFT, which indicate the topological sector to which the quasiparticles belong. Although the quasiparticle charges of the full theory are computed from the charge sector, the quantization conditions on the fundamental quasiparticle charges may be affected by the neutral sector. The edge structure of the monolayer anti-Pfaffian state is given by the following edge Lagrangian density,
\begin{align}
\mathcal{L}_{e} &= -\frac{\hbar}{4\pi} \sum_{ij} \left[ K_{ij} \partial_t \phi_i \partial_x\phi_i + V_{ij} \partial_x\phi_i\partial_x\phi_j \right] \nonumber \\
& ~~~+\frac{i\hbar}{4\pi} \psi \left( \partial_t - v_n\partial_x \right)\psi,
\end{align}
where $\phi_i$ represent the chiral Bose $U(1)$ fields belonging to the charge sector, and $\psi$ represent the neutral Majorana edge mode. The matrix $V_{ij}$ represents the inter-mode interaction and velocities of each mode. The propagation direction of the charge mode is to the right, thus the Majorana mode is upstream. The direction of the neutral Bose modes is set by the ${\bf K}$ matrix.

With this, we construct a monolayer ${\bf K}$-matrix for the anti-Pfaffian state. 
{\color{black} We closely follow the analysis from Ref. \onlinecite{hutasoit2017enigma}.}
Note that since ${}^2_0$CF lacks interlayer flux, a bilayer ${\bf K}$-matrix describing the $\nu_{\text{total}}=3/8+3/8$ state is simply a block diagonal matrix of the monolayer $\nu=3/8$ state. In the composite fermion picture, the state can be described as a daughter state of two $\Lambda$ levels completely filled with CF-electrons and half of one of the $\Lambda$ levels is filled with CF-holes. This results in a $3\times 3$ ${\bf K}_{\text{CF}}$-matrix of composite fermions. We then attach $2$ flux quanta to obtain the physical ${\bf K}$-matrix as ${\bf K} = {\bf K}_{\text{CF}}+2{\bf J}_{3}$, resulting in,
\begin{align}
{\bf K} = \begin{pmatrix}
3 & 2 & 2 \\
2 & 3 & 2 \\
2 & 2 & 0
\end{pmatrix}, ~~~~\text{with}~~~~{\bf t}=\begin{pmatrix}
1\\1\\1
\end{pmatrix},
\end{align} 
where ${\bf J}_{3}$ is a $3\times 3$ constant matrix of ones. The ${\bf K}$-matrix defines the $U(1)$ charged sector. We tensor this with a neutral anti-chiral part of the Ising CFT, which together defines the edge state structure of the anti-Pfaffian state \cite{hutasoit2017enigma,Mukherjee2012Pfaffian}. There are four chiral modes on the edge: two downstream bosonic modes, and two upstream, a bosonic and a Majorana mode. The edge structure determines the thermal Hall conductance \textcolor{black}{that can be straightforwardly extended to bilayer systems from the monolayer case \cite{hutasoit2017enigma}}.

We now look at the quasiparticle content of the theory. The operators in the Ising CFT are divided into three topological classes each of which are represented by the primary operators in the CFT \cite{hansson2017quantum}. The primary operators of the Ising CFT are, ${\bf 1}$ (Vacuum sector), $\psi$ (Fermion sector), and $\sigma$ (Ising vortex sector). The non-Abelian nature of the state can be seen from the non-trivial fusion rules computed via the operator product expansions of the primary fields,
\begin{align}
\psi\times\psi = {\bf 1},~~~\psi\times\sigma=\sigma,~~~\sigma\times\sigma={\bf 1}+\psi,
\end{align}
along with the trivial rule that all operators fuse trivially with the vacuum sector. One now obtains the spectrum of the theory by first identifying all the electron operators. The quasiparticle operators are then identified by requiring single-valued exchange with the electron operators. The \textcolor{black}{monolayer} quasiparticle operator for the fundamental charge is given by 
\begin{align}
\label{fund-eq}
\psi_{\text{fund.}}(x,t) = \sigma(x,t) ~\otimes e^{i\left[\phi_1(x,t)+3\phi_3(x,t)/2\right]},
\end{align} 
that has a quasiparticle charge of $e/16$, and a scaling dimension of 29/192, and belongs to the Ising vortex sector \cite{hutasoit2017enigma}. Here $\otimes$ represents the tensor product. We also denote the $U(1)$ bosonic modes as $\phi_i$ where $i=1,2,3$. In the bilayer state, a corresponding quasihole in the second layer is created with a charge $-e/16$. Although the quasiparticle charge only depends on the charge sector, the contribution to the self-statistical phase comes from the charge (Abelian), and neutral (non-Abelian) sectors.
{\color{black} The Abelian part of the phase is computed as $\pi {\bf l}^T {\bf K}^{-1} {\bf l}$ \textcolor{black}{$\mod 2\pi$}, where ${\bf l}^T=(1~0~3/2)$ can be read out from Eq. (\ref{fund-eq}).}  
The self-statistical (Abelian) phase for the fundamental quasiparticle on each layer is $\theta_s = -5\pi/32$ while the self-statistical (Abelian) phase of a $Q_{\text{exciton}}=\pm e/16$ exciton with itself is $\theta_s= -10\pi/32$. Note that these Abelian parts of the phases are accompanied by the non-Abelian contribution depending upon the fusion channels. 

Another property of interest is the thermal Hall conductance. Since the edge state structure of the anti-Pfaffian phase differs from the other Abelian and non-Abelian candidate states, thermal Hall conductance measurements provide important information about the ground state topological order. Ignoring edge reconstruction, in the units of the quantum of heat conductance $\kappa = (\pi^2k_B^2/3h)T$, each bosonic $U(1)$ mode carries a single unit of heat, while a Majorana mode contributes half a unit of heat \textcolor{black}{quanta \cite{affleck1988universal,blote1986conformal}}. Therefore, the edge state structure of a monolayer $\nu=3/8$ state should have the thermal Hall conductivity of $\kappa/2$, and similarly the observed bilayer state at $\nu_{\text{total}}=3/8+3/8$ has the thermal Hall conductivity of $\kappa = (\pi^2 k_B^2/3h)T$.

\newpage
\clearpage

\pagebreak
\begin{widetext}
\section{Supplementary Information}

\begin{center}
\textbf{\large Bilayer Excitons in the Laughlin Fractional Quantum Hall State}\\
\vspace{10pt}

Ron Q. Nguyen$^{\ast}$, Naiyuan James Zhang$^{\ast}$,
 Navketan Khurana-Batra$^{\ast}$, Sarah Alkidim, Xiaoxue Liu, 
 Kenji Watanabe, Takashi Taniguchi, D. E. Feldman, and J.I.A. Li$^{\dag}$

\vspace{10pt}
$^{\dag}$ Corresponding author. Email: jia$\_$li @brown.edu
\end{center}

\noindent\textbf{This PDF file includes:}

%\newpage
\renewcommand{\vec}[1]{\boldsymbol{#1}}
\renewcommand{\thefigure}{S\arabic{figure}}
\def\theequation{S\arabic{equation}}
\def\thetable{S\Roman{table}}
\setcounter{figure}{0}
\setcounter{equation}{0}

\noindent{Supplementary Text}\\
\noindent{Supplementary Data}

\section{Supplementary Text}

\subsection{I. Fractional charge and statistics of excitons belonging to ${}^2_0$CF family}

In this section we find the fractional charge and self-statistical phase for the Abelian interlayer excitons considered in the main text. We limit to the case when the effective filling $\nu_i^*\in \mathbb{Z}$ for each layer $i=1,2$. Since there are no interlayer fluxes, the bilayer ${\bf K}$-matrix is block diagonal ${\bf K}=\text{diag}({\bf K}_1,{\bf K}_2)$ with each block ${\bf K}_i$ being the monolayer ${\bf K}$-matrix along with their corresponding charge vector ${\bf t}_i$. Working in the symmetric representation ${\bf t}=({\bf t}_1\oplus{\bf t}_2)^T=(1,\dots,1)$, each monolayer ${\bf K}$-matrix takes a simple form,
\begin{align}
{\bf K}_i = \sigma_i{\bf I} + 2{\bf J} ~~~~\Leftrightarrow~~~~ {\bf K}^{-1}_i = \sigma_i{\bf I} - \frac{2\sigma_i}{2\dim{\bf K}_i+\sigma_i}{\bf J}~~~~~~~\text{for}~~i=1,2,
\end{align}  
where $\sigma_i=\pm 1$. The matrices ${\bf I}$ and ${\bf J}$ are identity and a constant matrix of ones respectively. Since $\dim {\bf K}_i$ represents the number of $U(1)$ gauge fields in the low energy Chern-Simons description, we can set $\dim {\bf K}_i=|\nu_i^*|$. Note that this identification only works for the Abelian states when $\nu_i^*\in \mathbb{Z}$ which we focus on in this section. Similarly, $\sigma_i$ is chosen depending upon the ${}^2_0$CF particle-/hole-doped $\Lambda$ levels so we identify $\sigma_i=\text{sign}(\nu_i^*)$. 

Next, we compute the fractional charge for an arbitrary excitation labeled by ${\bf l}_i$ in the $i$th layer,
\begin{align}
Q_i = -e ~{\bf t}^T_i {\bf K}_i^{-1} {\bf l}_i &= -e\sum_{n,m}^{|\nu_i^*|} \left( \sigma_i \delta_{nm} - \frac{2\sigma_i}{2|\nu_i^*|+\sigma_i} J_{nm} \right) \left[ {\bf l}_i\right]_m = -e \left( \sigma_i - \frac{2\sigma_i}{2|\nu_i^*|+\sigma_i} |\nu_i|^* \right) \sum_{m}^{|\nu_i^*|}  \left[ {\bf l}_i\right]_m \\
& = -e \frac{\ell_i}{\sigma_i\left( 2\nu_i^*+1\right)} ,
\end{align}
where in the last equality we used $\sigma_i=\text{sign}(\nu_i^*)$ and defined the integer $\ell_i \equiv \sum_{m}^{|\nu_i^*|}  \left[ {\bf l}_i\right]_m$. Since the vector ${\bf l}_i$ has integer components, the minimal charge excitation in each layer is $Q_{\text{min.},i}=\pm e/|2\nu_i^*+1|$ which is simply the fundamental charge in the Jain states. Next, we find the minimal exciton charge for the bilayer system via the condition $Q_1+Q_2=0$ such that $Q_{1,2}\ne 0$. The minimal integer solutions of $\ell_1,\ell_2\in\mathbb{Z}$ are,
\begin{align}
\frac{\ell_1}{|2\nu_1^* +1|} + \frac{\ell_2}{|2\nu_2^* +1|} = 0 ~~~~\Rightarrow ~~~~ \ell_i = \frac{(-1)^i ~|2\nu_i^*+1|}{\gcd(|2\nu_i^*+1|,|2\nu_2^*+1|)}~~~i=1,2,
\end{align}
where the integer valued function $\text{gcd}(n_1,n_2)$ computes the greatest common divisor of the integers $n_1$ and $n_2$. We also use $\text{sign}(\nu_i^*)=\text{sign}(2\nu_i^*+1)$ since $|\nu_i^*|\geq 1$. This gives the minimal exciton charge,
\begin{align}
Q_{\text{exciton}} = \pm \frac{e}{\gcd(|2\nu_i^*+1|,|2\nu_2^*+1|)}.
\end{align}
The minimal solutions $\ell_i$ can also be used to compute the exciton statistics. Before we compute the exciton statistics, we notice that due to the definition $\ell_i \equiv \sum_{m}^{|\nu_i^*|}  \left[ {\bf l}_i\right]_m$, there is an ambiguity in defining the excitation corresponding to the integer vector ${\bf l}_i$. This ambiguity suggests that there are distinct solutions of ${\bf l}$ that all correspond to the same exciton charge, or for that matter, a generic quasiparticle charge. However, to make sure all these distinct solutions describe the same particle, we need to check if all such excitations result in the same mutual and self statistical phases modulo $2\pi$. We proved that the distinct solutions indeed result in the same statistical phases in our previous report (see Ref. \cite{Zhang2025fractionalexciton}) for a general bilayer ${}^a_b$CF family. This guarantees that for any choice of ${\bf l}_i$ corresponding to the solutions obtained $\ell_i$ indeed describes the same minimal exciton as with any other choice of ${\bf l}_i$ preserving $\ell_i$. We thus set $\left[{\bf l}_1\right]_1 = \ell_1$ and $\left[{\bf l}_2\right]_1 = \ell_2$, and all the remaining elements are set to zero $[{\bf l}_i]_n=0$ $\forall ~n\in [2,|\nu_i^*|]$. The exciton self-statistics is computed as,
\begin{align}
\theta_s &= \pi \sum_{i=1}^{2} {\bf l}_i^T {\bf K}_i^{-1} {\bf l}_i \mod 2\pi = \pi \left[ {\bf K}_1^{-1} \right]_{11} \ell_{1}^2 + \pi \left[ {\bf K}_2^{-1} \right]_{11} \ell_{2}^2 \mod 2\pi \\
& = \frac{\pi}{\gcd(|2\nu_i^*+1|,|2\nu_2^*+1|)^2} \sum_{i=1}^{2}\left[ \sigma_i - \frac{2}{(2\nu_i^*+1)}\right] (2\nu_i^*+1)^2 \mod 2\pi\\
&= \pi \sum_{i=1}^{2} \frac{ \left( |2\nu_i^*+1|-2\right)\left(2\nu_i^*+1\right) }{\text{gcd}(|2\nu_1^*+1|,|2\nu_2^*+1|)^2} \mod 2\pi
\end{align}  
which shows that the excitons, in general, follow anyonic statistics rather than the usually anticipated bosonic character. Interestingly, we note that since $\nu_i^*\in \mathbb{Z}$, the sum $2x\equiv \sum_{i=1}^2 \left( |2\nu_i^*+1|-2\right)\left(2\nu_i^*+1\right) \in 2\mathbb{Z}$ for some integer $x\in \mathbb{Z}$. Similarly, the denominator $2y+1 \equiv \text{gcd}(|2\nu_1^*+1|,|2\nu_2^*+1|)^2 \in 2\mathbb{Z}+1$ for some $y\in \mathbb{Z}$. Now, since the following equation has no integer solutions,
\begin{align}
    \frac{x}{2y+1} \mod 1 = \frac{1}{2}~~~~\Rightarrow ~~~~ \text{no solution for $x,y\in \mathbb{Z}$},
\end{align}
the excitons in the above construction are never fermionic for any combination of monolayer Jain states $\nu_i^*\in \mathbb{Z}$. 

\textcolor{black}{\subsection{II. Drag current in the generalized (111) condensate phase}}

\textcolor{black}{We follow the previous analysis (see Supplementary Information in \cite{Zhang2025fractionalexciton}) to compute the drag current arising solely from the condensate phase. A few clarifications are in order. By `drag current due to the condensate phase', we mean the {\color{black}contribution generated by the backflow of the condensed ground state}. This backflow can be induced by interlayer flux attachments, which induce electric fields in the drag layer. In this situation, a perfect drag response (i.e., $j_{\text{drag}}=-j_{\text{drive}}$) is not guaranteed and will depend on the ground state, and our goal in this section is to quantify precisely this deviation. In contrast, a distinct mechanism, arising from quasiparticle/quasihole pairs forming an exciton between the two layers, yields a perfect drag response {\color{black} whenever excitons move as a whole and do not separate into independent opposite charges}. This latter mechanism is not the focus of this section.}

\textcolor{black}{Our starting point is the ${\bf K}$ matrix given in Eq. (\ref{eq:Kmat_111generalization}) in Methods. One way to interpret the ${\bf K}$ matrix is to treat the four $\Lambda$ levels as separate layers, each with different Hall response and conductance. Our goal is to compute the current density in drag layer $j_{\text{drag}}$ as a function of current density in drive layer $j_{\text{drive}}$. We apply an external electric field along $x$ direction in the drive layer, that should be the same in two $\Lambda$ levels corresponding to the drive layer. \textcolor{black}{The  perpendicular external magnetic field is assumed to be pointing in the $z$ direction, and therefore the attached fluxes point along negative $z$ direction.}}

\textcolor{black}{Since the induced electric field due to the motion of fluxes is ${\bf E}_{\text{ind}}=-{\bf v}\times {\bf B}$, the induced electric field in the $i$th layer is, $\delta E_{ix} = - \sigma_0^{-1} p_{ij} j_{jy} $ where $\sigma_0 = e^2/h$ is the conductance quantum. The interlayer and intralayer fluxes are denoted by $p_{ij}$. Similarly, the $y$ component of the induced electric field is given by, $\delta E_{iy} = \textcolor{black}{\sigma_0^{-1}} p_{ij} j_{jx}$. For the state described by ${\bf K}$-matrix Eq. (\ref{eq:Kmat_111generalization}) the flux attachments are $p_{ii}=2$ for all layers, and the interlayer fluxes are $p_{12}=p_{34}=2$ and $p_{23}=1$. }

\textcolor{black}{We now compute the current density in each layer due to the external applied electric field and the contributions form the induced electric field. We keep the longitudinal conductance $\sigma^i_{xx}\ne 0$ and assume rotational symmetry, thus $\sigma^i_{xx}=\sigma^i_{yy}$, and $\sigma^i_{xy}=\textcolor{black}{-\sigma^i_{yx}}$, for all $i=1,\dots,4$, we get,
\begin{subequations}
\begin{align}
j_{1x} &= \sigma^{1}_{xx} \left[ E - 2\sigma_0^{-1} \left( j_{1y} + j_{2y} \right) \right] - 2\sigma_{xy}^{1}\sigma_0^{-1} \left( j_{1x} + j_{2x} \right), \\
j_{2x} &= \sigma^{2}_{xx} \left[ E - \sigma_0^{-1} \left( 2 j_{1y} + 2 j_{2y} + j_{3y} \right) \right] - \sigma_{xy}^{2}\sigma_0^{-1} \left( 2 j_{1x} + 2 j_{2x} +  j_{3x} \right), \\
j_{3x} &= -\sigma^{3}_{xx} \sigma_0^{-1} \left(   j_{2y} + 2 j_{3y} + 2 j_{4y} \right) - \sigma_{xy}^{3}\sigma_0^{-1} \left( j_{2x} + 2 j_{3x} + 2 j_{4x} \right), \\
j_{4x} &= - 2\sigma^{4}_{xx} \sigma_0^{-1}  \left(   j_{3y} +  j_{4y} \right) - 2\sigma_{xy}^{4}\sigma_0^{-1} \left( j_{3x} +  j_{4x} \right),\\
j_{1y} &= 2\sigma_{xx}^{1} \sigma_0^{-1} \left( j_{1x} + j_{2x} \right) + \sigma_{xy}^{1} \left[ E - 2\sigma_0^{-1} \left( j_{1y} + j_{2y}  \right) \right],\\
j_{2y} &= \sigma_{xx}^{2} \sigma_0^{-1} \left( 2 j_{1x} + 2 j_{2x} +  j_{3x} \right) + \sigma_{xy}^{2} \left[ E - \sigma_0^{-1} \left(2 j_{1y} + 2 j_{2y} +  j_{3y}  \right) \right],\\
j_{3y} &= \sigma_{xx}^{3} \sigma_0^{-1} \left(  j_{2x} + 2 j_{3x} + 2 j_{4x} \right) - \sigma_{xy}^{3} \sigma_0^{-1} \left( j_{2y} + 2 j_{3y} + 2 j_{4y} \right), \\
j_{4y} &= 2\sigma_{xx}^{4} \sigma_0^{-1} \left(  j_{3x} + j_{4x} \right) - 2\sigma_{xy}^{4} \sigma_0^{-1} \left( j_{3y} + j_{4y}\right).
\end{align}
\end{subequations}
The drive and drag currents are defined as $j_{\text{drive}}=j_{1x}+j_{2x}$ and $j_{\text{drag}}=j_{3x}+j_{4x}$. The above expressions can be simplified by substituting the quantized value of the Hall conductance of each layer as $\sigma^i_{xy}\sigma_{0}^{-1}\equiv \nu_i$, where $\nu_i=1$ for all $i$. Next, from Ohm's law $j_x\sim \sigma_{xx}E$, therefore, we can rewrite these coupled equations up to first order in $\alpha_i$, where we define $\alpha_{i}\equiv \sigma_{xx}^i \sigma_{0}^{-1}$.
\begin{subequations}
\begin{align}
j_{1x} &= \alpha_{1} \left( E\sigma_0 -  2 j_{1y} - 2 j_{2y} \right) -  2\left( j_{1x} +  j_{2x} \right), \\
j_{2x} &= \alpha_2 \left( E\sigma_0 -  2 j_{1y} - 2 j_{2y} - j_{3y}  \right) - \left( 2 j_{1x} + 2 j_{2x} + j_{3x} \right),\\
j_{3x} &= -\alpha_3 \left(   j_{2y} +2 j_{3y} + 2 j_{4y} \right) -  \left(  j_{2x} + 2 j_{3x} + 2 j_{4x} \right), \\
j_{4x} &= - 2\alpha_4 \left(   j_{3y} +  j_{4y} \right) - 2\left(  j_{3x} +  j_{4x} \right),\\
j_{1y} &=  \left( E\sigma_0 - 2 j_{1y} - 2 j_{2y}  \right) + \mathcal{O}(\alpha_{1}^2), \\
j_{2y} &= \left( E\sigma_0 - 2 j_{1y} - 2 j_{2y} -  j_{3y}  \right)  + \mathcal{O}(\alpha_{2}^2), \\
j_{3y} &= - \left(  j_{2y} + 2 j_{3y} + 2 j_{4y} \right) + \mathcal{O}(\alpha_{3}^2), \\
j_{4y} &= -  2 \left(  j_{3y} + j_{4y}\right) + \mathcal{O}(\alpha_{4}^2).
\end{align}
\end{subequations}
The solution for the transverse current density is $j_{1y}=j_{4y}=E\sigma_{0}/8$, $j_{2y}=5E\sigma_{0}/16$, and $j_{3y}=-3E\sigma_{0}/16$. Plugging these into the expressions for longitudinal current, we obtain,
\begin{subequations}
\begin{align}
j_{1x}&=\alpha_1 \frac{E\sigma_0}{8} -2 \left( j_{1x} + j_{2x} \right),\\
j_{2x}&=\alpha_2 \frac{5 E\sigma_0}{16} - \left( 2j_{1x} + 2j_{2x} + j_{3x} \right),\\
j_{3x} &= -\alpha_3 \frac{3E\sigma_0}{16} - \left( j_{2x} + 2j_{3x} + 2j_{4x} \right),\\
j_{4x} &= \alpha_4 \frac{E\sigma_0}{8} - 2 \left( j_{3x} + j_{4x} \right).
\end{align}
\end{subequations}
The above equations can now be solved for the longitudinal current. The solution is,
\begin{subequations}
\begin{align}
j_{1x} =& \frac{3}{32}\sigma^1_{xx}E - \frac{25}{128}\sigma_{xx}^2E - \frac{9}{128} \sigma_{xx}^3E - \frac{1}{32} \sigma^{4}_{xx}E, \\
j_{2x} =& -\frac{5}{64} \sigma_{xx}^1 E + \frac{75}{256} \sigma^2_{xx} E + \frac{27}{256}\sigma_{xx}^3 E +  \frac{3}{64} \sigma_{xx}^{4} E, \\
j_{3x} =& \frac{3}{64}\sigma_{xx}^1E - \frac{45}{256}\sigma_{xx}^2 E  -\frac{45}{256} \sigma_{xx}^3 E - \frac{5}{64} \sigma_{xx}^4 E,\\
j_{4x} =& -\frac{1}{32} \sigma_{xx}^1 E + \frac{15}{128} \sigma_{xx}^2 E + \frac{15 }{128} \sigma_{xx}^3E + \frac{3}{32} \sigma_{xx}^4 E.
\end{align}
\end{subequations}
This solution now can be used to find the longitudinal drag/drive currents given by the definition $j_{\text{drive}}\equiv j_{1x}+j_{2x}$ and $j_{\text{drag}}\equiv j_{3x}+j_{4x}$. We have,
\begin{subequations}
\begin{align}
j_{\text{drive}} &= \frac{1}{64} \sigma_{xx}^1 E + \frac{25}{256}\sigma_{xx}^2 E + \frac{9}{256} \sigma_{xx}^3 E + \frac{1}{64} \sigma_{xx}^4E,\\
j_{\text{drag}} &= \frac{1}{64}\sigma_{xx}^1 E - \frac{15}{256} \sigma_{xx}^2 E - \frac{15}{256} \sigma_{xx}^3 E + \frac{1}{64} \sigma_{xx}^4 E. 
\end{align}
\end{subequations}
We can now simplify this further by imposing that the conductance of the first and fourth layer is the same $\sigma_{xx}^1=\sigma_{xx}^4\equiv\sigma^{\text{I}}_{xx}$. Similarly, the second and third layer have the same conductance as well $\sigma_{xx}^2=\sigma_{xx}^3\equiv\sigma^{\text{II}}_{xx}$. We get,
\begin{subequations}
\begin{align}
j_{\text{drive}} &= \frac{4}{128} \sigma_{xx}^{\text{I}}E + \frac{17}{128} \sigma_{xx}^{\text{II}} E, \\
j_{\text{drag}} &= \frac{4}{128} \sigma_{xx}^{\text{I}}E - \frac{15}{128} \sigma_{xx}^{\text{II}}E.
\end{align}
\end{subequations}
One cannot explicitly assume that the longitudinal conductance of the two $\Lambda$-levels to be the same. The drag ratio is,
\begin{align}
    \frac{j_{\text{drag}}}{j_{\text{drive}}} = \frac{4\sigma_{xx}^{\text{I}}-15\sigma_{xx}^{\text{II}}}{4\sigma_{xx}^{\text{I}}+17\sigma_{xx}^{\text{II}}}.
\end{align}
Physically at low temperatures, we expect $\sigma^{\text{I}}_{xx}\ll \sigma^{\text{II}}_{xx}$. However, the state described by Eq. (\ref{eq:Kmat_111generalization}) does not exhibit perfect drag, therefore, one cannot attribute the perfect drag response solely due to the condensate.}

\end{widetext}

\renewcommand{\thefigure}{S\arabic{figure}}

%\begin{figure*}
%\includegraphics[width=0.6\linewidth]{FigActivationEnergyv2.pdf}
%\caption{\label{Arrhenius}{\bf{Activated behavior of quasiparticle excitations.}}  (a,b) Arrhenius plot of bulk conductance measured in the parallel flow $G_{\text{PF}}$ (black) and counterflow $G_{\text{counter}}$ (blue) configurations at $\nu_{tot}=2/3$ and $B = 28$~T for a) $\Delta \nu=0$ and b) $\Delta \nu=0.06$. The activated behavior of $G_{\text{PF}}$ is directly linked to the charge gap, $\Delta_{\text{charge}}$, which represents the energy required to create thermally-excited quasiparticles and quasiholes. The activated behavior of $G_{\text{counter}}$ reveals the energy required to generate a charge-neutral excitation, $\Delta_{\text{exciton}}$. In the presence of a robust charge gap, bulk conductance in the counterflow geometries show activated behavior for thermally-generated excitons but little to no activation for doping-induced excitons.}
%\end{figure*}

\begin{figure*}
\includegraphics[width=0.9\linewidth]{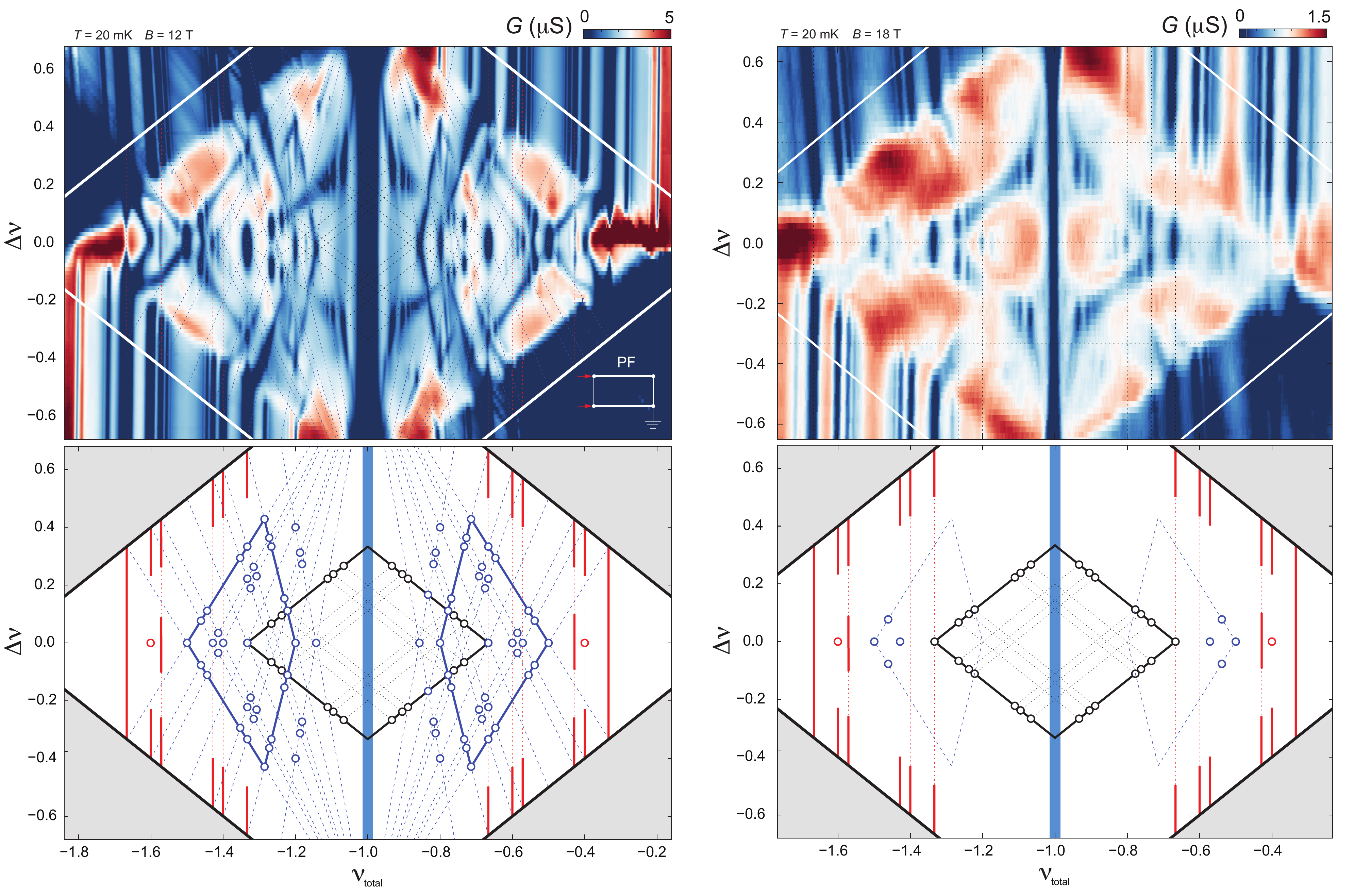}
\caption{\label{map_evolution}{\bf{Evolution of FQH states across $\nu_{total}-\Delta \nu$ map.}} Parallel-flow conductance $G_\mathrm{PF}$ measured in the Corbino device as functions of total filling factor $\nu_\mathrm{total}$ and $\Delta \nu$ at $B=12$~T (left) and $18$ T (right). The schematic diagram in the bottom panels highlight the observed FQH states at each $B$ field. Composite fermions with 2+1 flux attachment dominates at $B = 12$ T, whereas the Jain sequence states dominate at $B = 18$ T. 
}
\end{figure*}

% Schematic phase diagram of a quantum Hall bilayer, parameterized by $\nu_{\mathrm{total}}$ and $\Delta\nu$
\begin{figure*}
\includegraphics[width=0.8\linewidth]{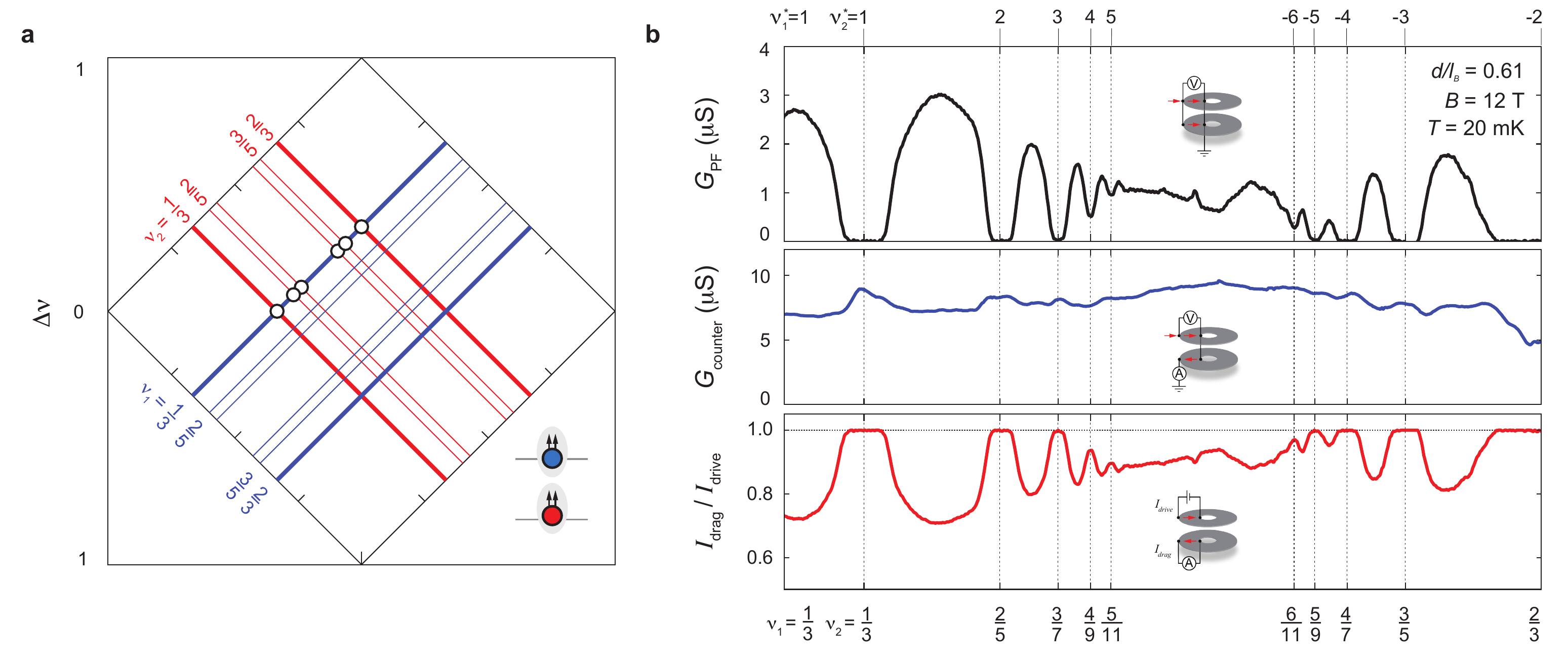}
\caption{\textbf{Excitonic pairing between single-component FQH states.}
(a) Schematic phase diagram of a quantum Hall bilayer as a function of $\nu_{\mathrm{total}}$ and $\Delta\nu$. Red and blue lines show the the expected trajectories for the single-component FQH states of the top and bottom layers, respectively. (b) Parallel-flow conductance $G_{\mathrm{PF}}$ (top), counterflow conductance $G_{\mathrm{counter}}$ (middle), and drag ratio $I_{\mathrm{drag}}/I_{\mathrm{drive}}$ (bottom) measured at fixed $\nu_1 = 1/3$ as functions of $\nu_2$ for $B = 12$~T and $T = 20$~mK in the Corbino device. This measurement corresponds to the same sweep shown in Fig.~3c, but acquired at a lower temperature. 
}
\end{figure*}

\begin{figure*}
\includegraphics[width=0.5\linewidth]{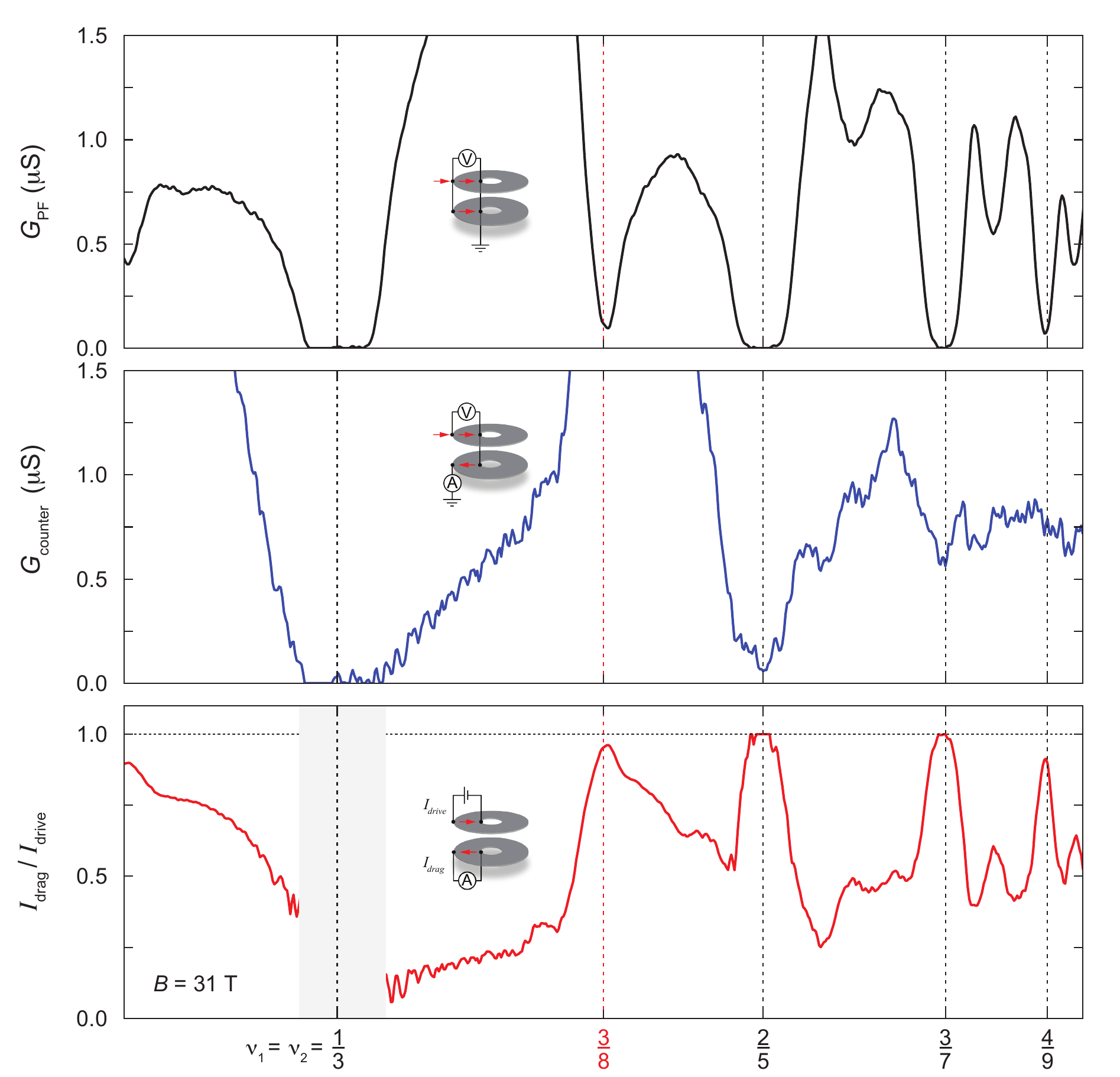}
\caption{\label{zoomed_EDL}\textbf{Equal density line in the Corbino device}.
Parallel-flow conductance $G_\mathrm{PF}$ (top), counterflow conductance $G_\mathrm{counter}$ (middle), and drag ratio $I_\mathrm{drag}/I_\mathrm{drive}$ (bottom) measured as functions of the total filling factor $\nu_\mathrm{total}$ at $B = 31$~T along the equal-density line. This dataset is identical to that shown in Fig.~4b--d, but here we adjust the vertical scales of $G_\mathrm{PF}$ and $G_\mathrm{counter}$ to highlight their distinct behaviors, which together give rise to the observed perfect drag response.
}
\end{figure*}

\begin{figure*}
\includegraphics[width=0.35\linewidth]{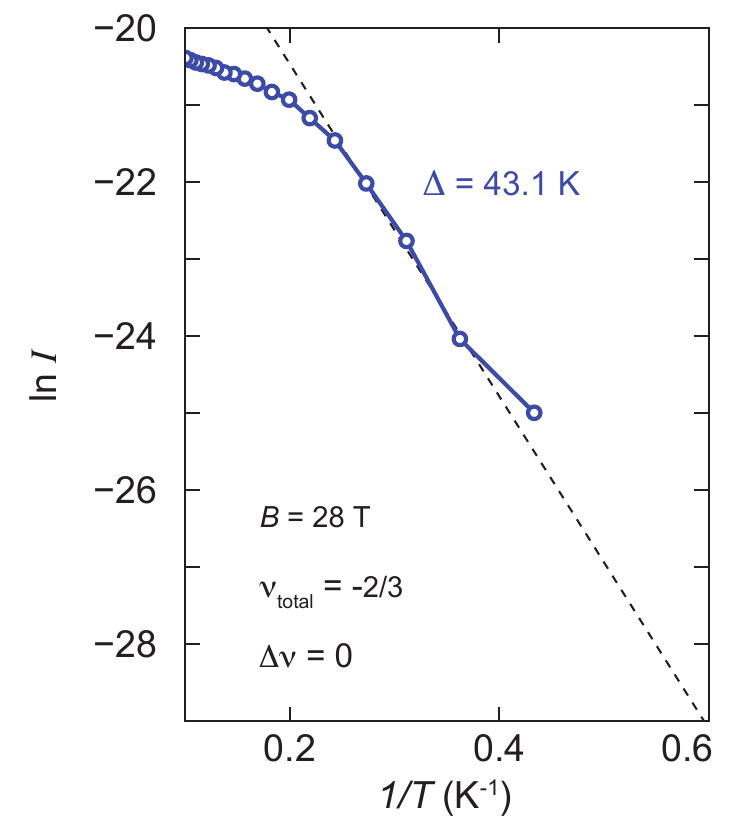}
\caption{\label{CFactivate}\textbf{Activated behavior in counterflow conductance.} 
Arrhenius plot of the counterflow conductance from the Corbino device as a function of temperature at $\nu_{\mathrm{total}} = -2/3$, $\Delta\nu = 0$, and $B = 28$~T, demonstrating activated behavior. We note that both unpaired charges and excitons contribute to the counterflow conductance. As a result, the observed activation behavior does not directly reflect the energy cost associated with generating excitons.
}
\end{figure*}

\end{document}